\documentclass[aps,prc,floatfix,twocolumn]{revtex4}

\usepackage{epsfig}
\usepackage{amsmath}

\begin{document}

\title{Interference of thermal photons from quark and hadronic phases in relativistic collisions of heavy nuclei}

\author{Dinesh Kumar Srivastava and Rupa Chatterjee}

\affiliation{Variable Energy Cyclotron 
Centre, 1/AF Bidhan Nagar, Kolkata 700 064, India}            
\date{\today}

\begin{abstract}
We explore the intensity correlations for thermal photons 
having $K_T\leq$ 2 GeV/$c$, for central collisions of 
heavy nuclei at RHIC and LHC energies. These photons  
get competing contributions from the quark and the hadronic 
phases. This competition gives rise to a unique structure, especially in 
the outward correlation function, due to the interference 
between the photons from the two sources. The temporal 
separation of the two sources provides the life time of the 
system and their strengths provide the relative contribution 
of the two phases. The results are found to be quite sensitive 
to the quark-hadron phase transition temperature and the formation
time of the plasma. 
\end{abstract}

\maketitle
\section{Introduction}
The last two decades have witnessed a concerted and a
well co-ordinated  theoretical and experimental effort 
to produce and study quark-gluon plasma - the deconfined 
strongly interacting matter, in relativistic collisions 
of heavy nuclei. The eminent commissioning of the Large 
Hadron Collider (LHC) at CERN and various upgrades, both 
in the accelerator and the detection systems, at the 
Relativistic Heavy Ion Collider (RHIC) at  Brookhaven will 
offer further opportunities  to advance our understanding 
of the physics of this novel state of matter, which 
permeated the early universe. We have already been rewarded 
with the discovery of the jet-quenching~\cite{jet1,jet2}, 
the elliptic flow~\cite{fl1,fl2}, and  the successful 
formulation of the partonic recombination~\cite{rec} as 
a model for hadronization in recent experiments at RHIC. 
The thermal photons radiated from such collisions are 
being studied to determine the high temperatures reached 
in these collisions~\cite{wa98_1,dks_wa98,phenix,david}.

The next step in this endeavour involves  a critical 
examination of our concepts about the formation and the 
evolution of the plasma. An important question in this 
connection is how quickly, if at all, does the plasma 
thermalize; so that we can use the powerful methods of 
hydrodynamics to model the evolution of the system. 
Theoretical estimates for the thermalization time ($\tau_0$)
vary considerably and experimental observables like elliptic 
flow of hadrons only suggest that $\tau_0 \sim 1.0$ fm/$c$. 
A more quantitative experimental determination of $\tau_0$ 
would be very valuable. One would also like to know the lifetime
of the interacting system. Can we determine the temperature at 
which the phase transition takes place?

The quantum statistical interference between identical 
particles emitted from these collisions is expected to provide 
valuable inputs for the space-time description of the system.
The use of photons for these studies admits several advantages.  
Firstly, they interact only weakly with the system after their 
production. Thus, they are not subjected to distorting effects 
such as re-scattering and Coulomb interactions which affect the 
results for hadrons. Secondly, and even more importantly, 
photons are emitted from every stage of the collision dynamics. 
These aspects give us a hope of getting a direct information about 
the earliest, hot and dense stage of the system by studying the 
photons having larger transverse momenta, $K_T$. Recently, some 
additional sources of large $K_T$ photons have also been 
proposed~\cite{bms_phot, fms_phot}. 

The difficulty, of-course, arises from a meager emission of 
direct photons which lie buried in  the huge background 
of decay photons. So far only one measurement involving 
photons having very low $K_T$ by the WA98 Collaboration~\cite{wa98_2} 
for the central collision of lead nuclei at CERN SPS has 
been reported.

On the other hand, the theory of the intensity interferometry 
of photons from relativistic heavy ion collisions has been 
pursued in considerable detail by several authors~\cite{dks1, dks2, ors}.
It is generally felt that the experimental efforts for these
studies  have a larger likelihood of success at RHIC and LHC 
energies because of larger initial temperature of the plasma  
and a large suppression of pions due to jet-quenching. Of late, 
there have also been tremendous advances in methods for 
identification of single photons~\cite{phenix}. 

In the present work, we focus  our attention on the photons 
having intermediate $K_T \approx $ 0.2 -- 2 GeV/$c$. The 
photons having $K_T \ll$ 2 GeV/$c$ will mostly originate 
from the hadronic phase of the system. They are expected to 
reveal a source which should be strongly affected by the flow 
and expansion of the system. The photons having $K_T \gg$ 2 
GeV/$c$ should unveil a source which is in the infancy of the 
hot and dense quark-gluon plasma, and where the flow has just 
started to develop. The photons having $K_T \leq$ 2 GeV/$c$ are 
unique. They have their origin either in the hot and dense quark 
phase of the system or in the relatively cooler but rapidly 
expanding hadronic phase, where a large build-up of the radial 
flow boosts their transverse momenta. 

We shall see that this leads to a rich structure in the 
correlation function for thermal photons especially when 
studied as a function of the outward momentum difference 
($q_o$, see later), due to the interference of the two 
sources. The two sources also manifest 
in the correlation functions for the longitudinal ($q_{\it l}$) 
and sideward ($q_s$) momentum differences. This renders the
correlation very sensitive to the formation time of the plasma. 
An early formation and thermalization would provide a large 
initial temperature, while a late formation and thermalization 
will lead to a smaller initial temperature. This analysis can 
also help to find the fractional contributions of the quark 
and the hadronic phases to the single photon spectrum.

Since the interference mentioned above is controlled by 
the relative contributions from the quark and the hadronic 
phases, which in turn are decided by the critical temperature
used in the model, we find a unique sensitivity of the 
results to the temperature at which the phase transition 
takes place.

We discuss the basic formalism in the next section. The 
results for RHIC and LHC energies are discussed in sect. 
III and IV respectively. In sect. V we discuss the  
sensitivity of our results to the initial formation time 
of the plasma and the transition temperature. Finally, we 
summarize our findings in sect. VI.

\section{Formulation}

One can define the spin averaged intensity correlation between 
two photons with momenta ${\bf k_1}$ and ${\bf k_2}$, emitted 
from a completely chaotic source, as:
\begin{equation}
C(\mathbf{q},\mathbf{K})=1+\frac{1}{2}
\frac{\left|\int \, d^4x \, S(x,\mathbf{K})
e^{ix \cdot q}\right|^2}
                        {\int\, d^4x \,S(x,\mathbf{k_1}) \,\, \int d^4x \,
S(x,\mathbf{k_2})}
\label{def}
\end{equation}
where $S(x,{\bf K})$ is the space-time emission function, and
\begin{equation}
\mathbf{q}=\mathbf{k_1}-\mathbf{k_2}, \,\, \,
 \mathbf{K}=(\mathbf{k_1}+\mathbf{k_2})/2\, \, .
\end{equation}
We shall use hydrodynamics to model the evolution of 
the system. The space-time emission function $S$ is
approximated as the rate of production of photons, 
$\rm{EdN/d^4x d^3k}$, from the quark and the hadronic 
phases.

The interference between the thermal photons 
from quark and the hadronic phases is best studied by 
writing the source function $S$ as $\rm {S_Q+S_H}$ in the 
numerator, where $Q$ and $H$ stand for the two phases 
respectively, and then including either one, or the other, 
or both.
 
We shall discuss the results for the correlation function
$C({\bf q},{\bf K})$ in terms of the outward, sideward,  
and longitudinal momentum differences, $q_{\text{o}}$, 
$q_{\text{s}}$ and $q_{\ell}$. Thus writing the 4-momentum 
as  $k_i^\mu$ of the ${\it{i}}$th photon, we have
\begin{equation}
k_i^\mu=(k_{iT}\, \cosh \, y_i,\mathbf{k_i})
\end{equation}
with
\begin{equation}
\mathbf{k_i}=(k_{iT}\,\cos \psi_i,k_{iT} \sin \psi_i,k_{iT} \, \sinh\, y_i),
\end{equation}
where $k_T$ is the transverse momentum, $y$ is the rapidity, 
and $\psi$ is the azimuthal angle. Defining the difference 
and the average of the transverse momenta,
\begin{equation}
\mathbf{q_T}=\mathbf{k_{1T}}-\mathbf{k_{2T}} \, \, ,
\mathbf{K_T}=(\mathbf{k_{1T}}+\mathbf{k_{2T}})/2\, ,
\end{equation}
we can write~\cite{dks1},
\begin{eqnarray}
q_{\ell}&=&k_{1z}-k_{2z}\nonumber\\
        &=&k_{1T} \sinh y_1 - k_{2T} \sinh y_2\\
\label{q_l}
q_{\text{o}}&=&\frac{\mathbf{q_T}\cdot \mathbf{K_T}}{K_T}\nonumber\\
       &=& \frac{(k_{1T}^2-k_{2T}^2)}
         {\sqrt{k_{1T}^2+k_{2T}^2+2 k_{1T} k_{2T} \cos (\psi_1-\psi_2)}}\\
\label{q_o}
q_{\text{s}}&=&\left|\mathbf{q_T}-q_{\text{o}}
       \frac{\mathbf{K_T}}{K_T}\right|\nonumber\\
        &=&\frac{2k_{1T}k_{2T}\sqrt{1-\cos^2(\psi_1-\psi_2)}}
      {\sqrt{k_{1T}^2+k_{2T}^2+2 k_{1T} k_{2T} \cos (\psi_1-\psi_2)}} \, .
\label{q_s}
\end{eqnarray}

The radii corresponding to the above momentum differences are 
often obtained by approximating the correlation function as,
\begin{eqnarray}
 C(q_{\text{o}}, q_{\text{s}},q_{\ell}) = 1 +  \frac{1}{2}\exp 
\left[-\left(q_{\text{o}}^2R_{\text{o}}^2 +  q_{\text{s}}^2R_{\text{s}}^2 
+ q_{\ell}^2R_{\ell}^2 \right ) \right].
\label{c1}
\end{eqnarray}
We also define the root mean square momentum difference $\langle 
q_i^2 \rangle$ as,
\begin{eqnarray}
\langle q_i^2 \rangle \, = \, \frac{\int \, (C-1)\, q_i^2 \, dq_i}{\int \, (C-1) \, dq_i},
\label{qi}
\end{eqnarray}
so that for the Gaussian parameterization given in Eq.~\ref{c1}, we have
\begin{eqnarray}
R_i^2 \, = \, \frac{1}{2 \, \langle q_i^2 \rangle} \, .
\end{eqnarray}
Thus, $1/[2\, \langle q_i^2 \rangle]^{1/2}$ becomes a useful measure when the 
correlation function has a more complex nature as we shall see later.

We consider central collision of gold  and lead nuclei, 
corresponding to the conditions realized at the RHIC 
and LHC, respectively. We assume that a thermally and 
chemically equilibrated quark-gluon plasma is produced 
at an initial time $\tau_0$. We further assume an 
isentropic expansion of the system to estimate the 
initial temperature, $T_0$, in terms of particle 
rapidity density. Thus,
\begin{equation}
\frac{2\pi^4}{45\zeta(3)} \frac{1}{A_T}\frac{dN}{dy}=4aT_0^3\tau_0 \, ,
\end{equation}
where $A_T$ is the transverse area of the system, ${\rm dN/dy}$ 
is the particle rapidity density, and $a=42.25 \pi^2/90$ for a 
plasma of massless quarks (u, d, and s) and gluons. The number 
of flavours for this purpose is taken as $\approx 2.5$ to 
account for the mass of the strange quarks. For the present 
work we shall mostly consider $\tau_0 \, = \, 0.2$ fm/$c$ and 
give illustrative results for $\tau_0$ varying from 0.2 to 
1.0 fm/$c$, keeping $\text{dN/dy}$ (or the total entropy) fixed. 
We add that, for study of thermal photons a value close to 0.2 
fm/$c$ may be more appropriate. The initial energy density is 
taken as a weighted sum of wounded nucleon and binary collision 
distributions as in earlier studies~\cite{puzzle}. The quark-hadron 
phase transition is assumed to take place at a temperature of 180 
MeV, while the freeze-out takes place at 100 MeV. The relevant 
hydrodynamic equations are solved under the assumption of 
boost-invariant longitudinal and azimuthally symmetric transverse 
expansion using the procedure discussed earlier~\cite{dks_hyd} and 
integration performed over the history of evolution.
We use the complete leading order results for the production 
of photons from the quark matter~\cite{guy}, and the results 
from Turbide {\it et al.}~\cite{simon} for radiation of photons 
from the hadronic matter. A rich equation of state with the 
inclusion of all the particles in the particle data book, 
having $M < $ 2.5 GeV/$c^2$ describes the hadronic matter. We 
take $\rm{dN/dy}$ at $y=0$ as 1260~\cite{fms_phot} for 200A 
GeV Au+Au collisions at RHIC and 5625~\cite{LHC} for 5.5A TeV 
Pb+Pb collisions at LHC. A smaller value for the particle 
rapidity density at LHC may perhaps be more appropriate, though. 
This will, however, not change the nature of the findings 
reported here.

\begin{figure}[t]
\centerline{\includegraphics[height=12.0cm,clip=true]{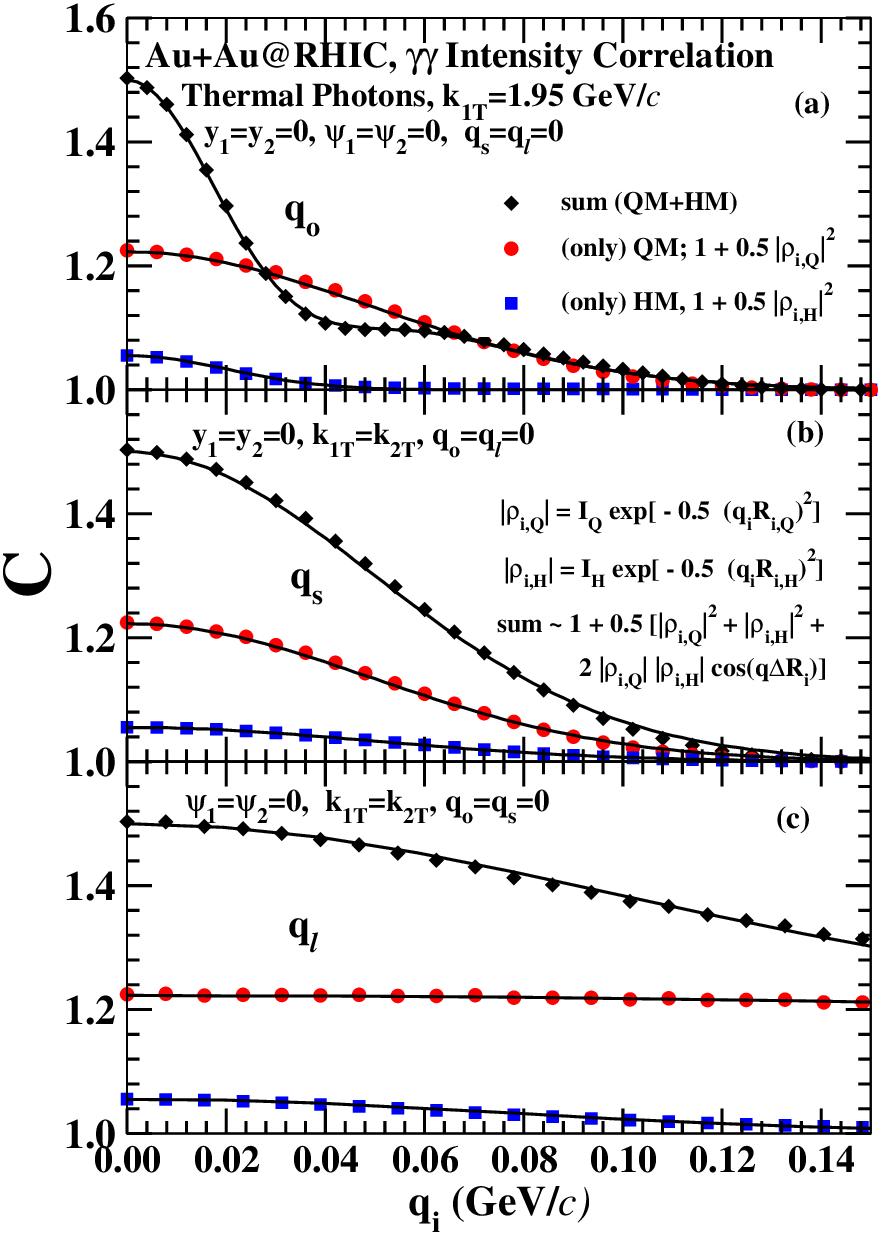}}
\caption{(Colour on-line) (a) The outward, (b) sideward, and 
(c) longitudinal correlation functions for thermal photons 
produced in central collision  of gold nuclei at RHIC taking 
$\tau_0 = 0.2$ fm/$c$. Symbols denote the results of the 
calculation, while the curves denote the fits.}
\label{fig1}
\end{figure}

\section{Results for RHIC energies}

As a first step, in Fig.~\ref{fig1} we have shown 
the outward, sideward, and longitudinal correlation 
functions for thermal photons at  
RHIC having $K_T\approx$ 2 GeV/$c$. The 4-momenta of the 
two photons are chosen so that when we study the outward 
correlations, the $q_{\text{s}}$ and $q_{\ell}$ are 
identically zero and the dependence on $q_{\text{o}}$ 
is clearly seen, and so on. We first discus the results 
when only the hadronic matter contribution or only the 
quark matter contribution is included in numerator 
(Eq.~\ref{def}). We find that the correlation functions 
for  the two phases can be approximated as:
\begin{eqnarray}
C(q_i, \alpha) = 1 \ + \ 0.5 |\rho_{i,\alpha}|^2 
\end{eqnarray}
where i = o, s, and $\ell$, and $\alpha$ denotes quark matter (Q) 
and hadronic matter (H) in an obvious notation. The source 
distribution $|\rho_{i,\alpha}|$ is very well described by,
\begin{eqnarray}
 |\rho_{i,\alpha}| \ = \ I_i \ \rm{exp}\ \left [- \ 0.5 \ 
(q_i^2R_{i,\alpha}^2) \right ] 
\end{eqnarray}
where, 
\begin{eqnarray}
I_Q &=& {\frac {dN_Q}{(dN_Q\,+\,dN_H)}},
\end{eqnarray}
and
\begin{eqnarray}
I_H &=& {\frac {dN_H}{(dN_Q\,+\,dN_H)}}. 
\end{eqnarray}
The final correlation function denoted by `sum' in the 
figures can be approximated as:
\begin{eqnarray}
C(q_i) \ = \  1 \ & + & \ 0.5 \left[ \ |\rho_{i,Q}|^2 \ +  \ 
|\rho_{i,H}|^2 \ \right.\nonumber\\
& + & \left. \ 2 \ |\rho_{i,Q}| |\rho_{i,H}| \ \rm{cos}(q\Delta R_i) 
\right] 
\label{cc} 
\end{eqnarray} 
which clearly brings out the interference between the two 
sources~\cite{yves}. Here $\Delta R_i$ stands for the 
separation of the two sources in space-time and $q$ is 
the 4-momentum difference. For thermal photons having 
$K_T \approx 2$ GeV/$c$ at RHIC, various radii (in fm) 
are obtained as:
\begin{eqnarray}
R_{o,Q}& = & 2.8, \ R_{o,H}  =  7.0,  \ \Delta R_o  = 12.3, \nonumber\\
R_{s,Q} & \approx\ & R_{s,H}  =  2.8, \ \Delta R_s  \approx 0., \nonumber\\
R_{\ell,Q} &= & 0.3, \ R_{\ell,H}  =  1.8, \ \Delta R_\ell  \approx  0.
\end{eqnarray}

These results imply~\cite{yves} that, while the spatial 
separation of the two sources is negligible, their temporal 
separation is about 12 fm. This gives the life-time of the 
system. If the mixed phase is of shorter duration or absent, 
this will obviously decrease.
%%%%%%%%%%%%%%%%%%%%%%%%%%%%%%%%%%%%%%%%%%
\begin{figure}[tb]
\includegraphics[height=5.60cm, clip=true]{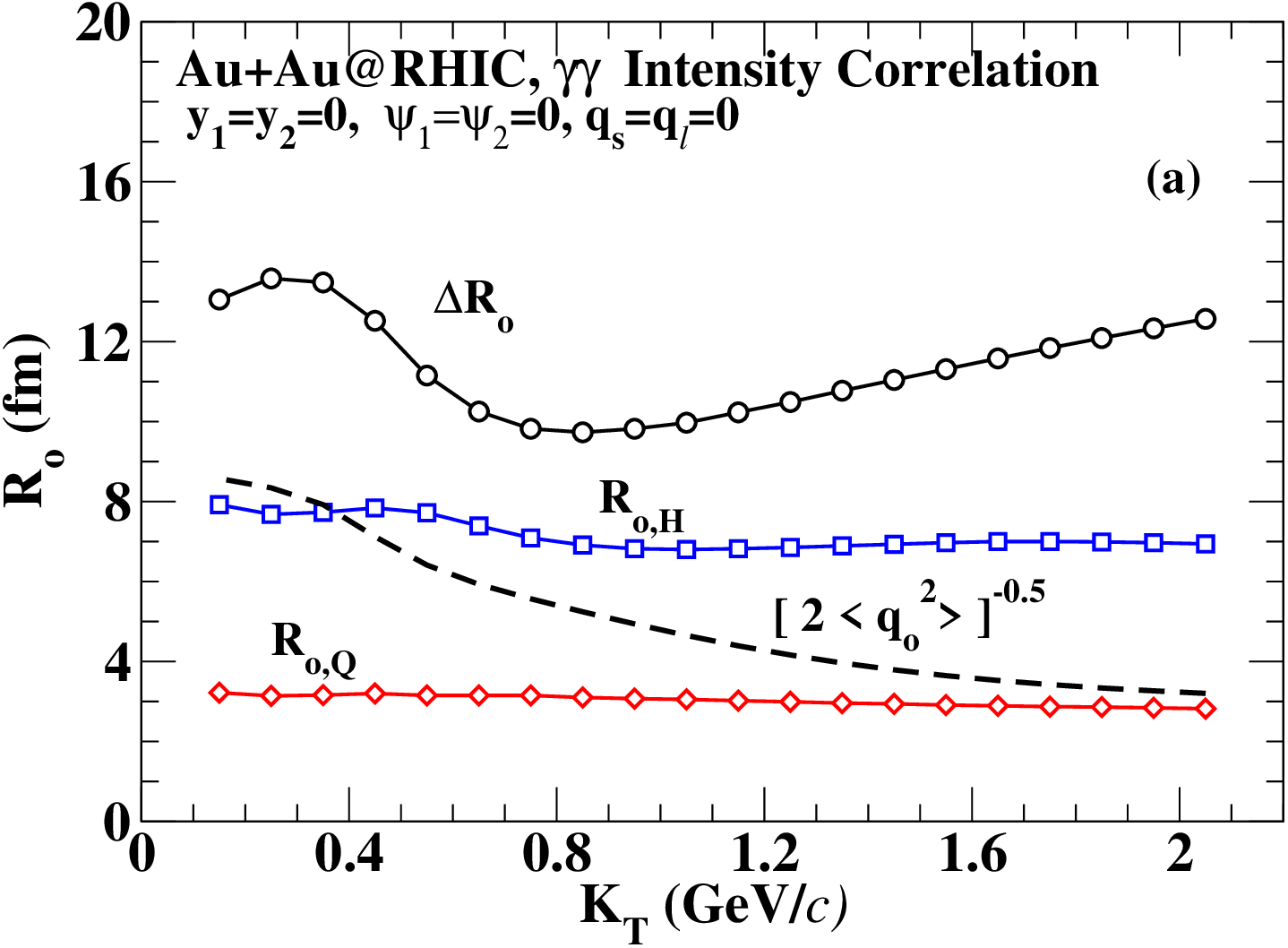}
\includegraphics[height=5.60cm, clip=true]{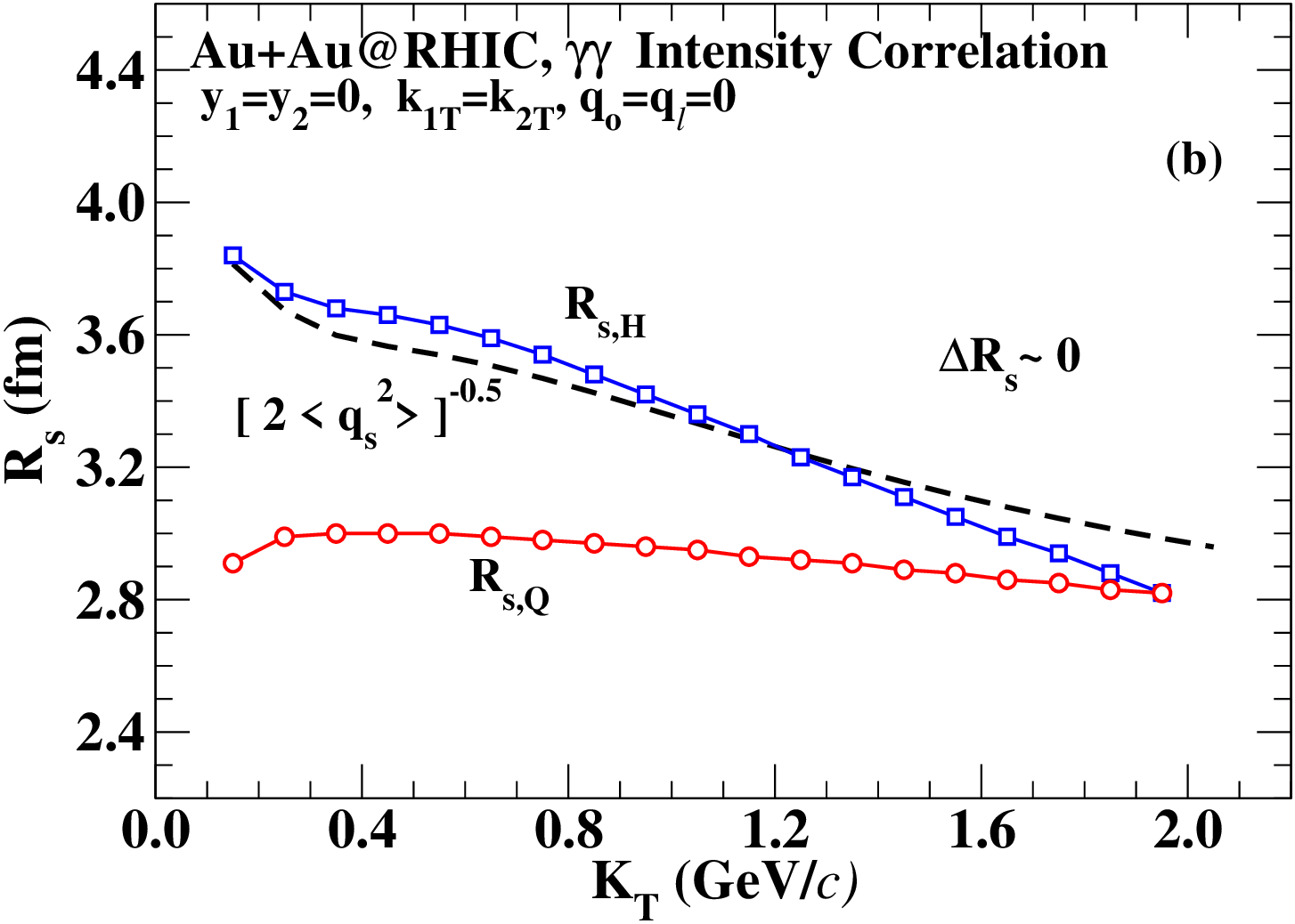}
\includegraphics[height=5.60cm, clip=true]{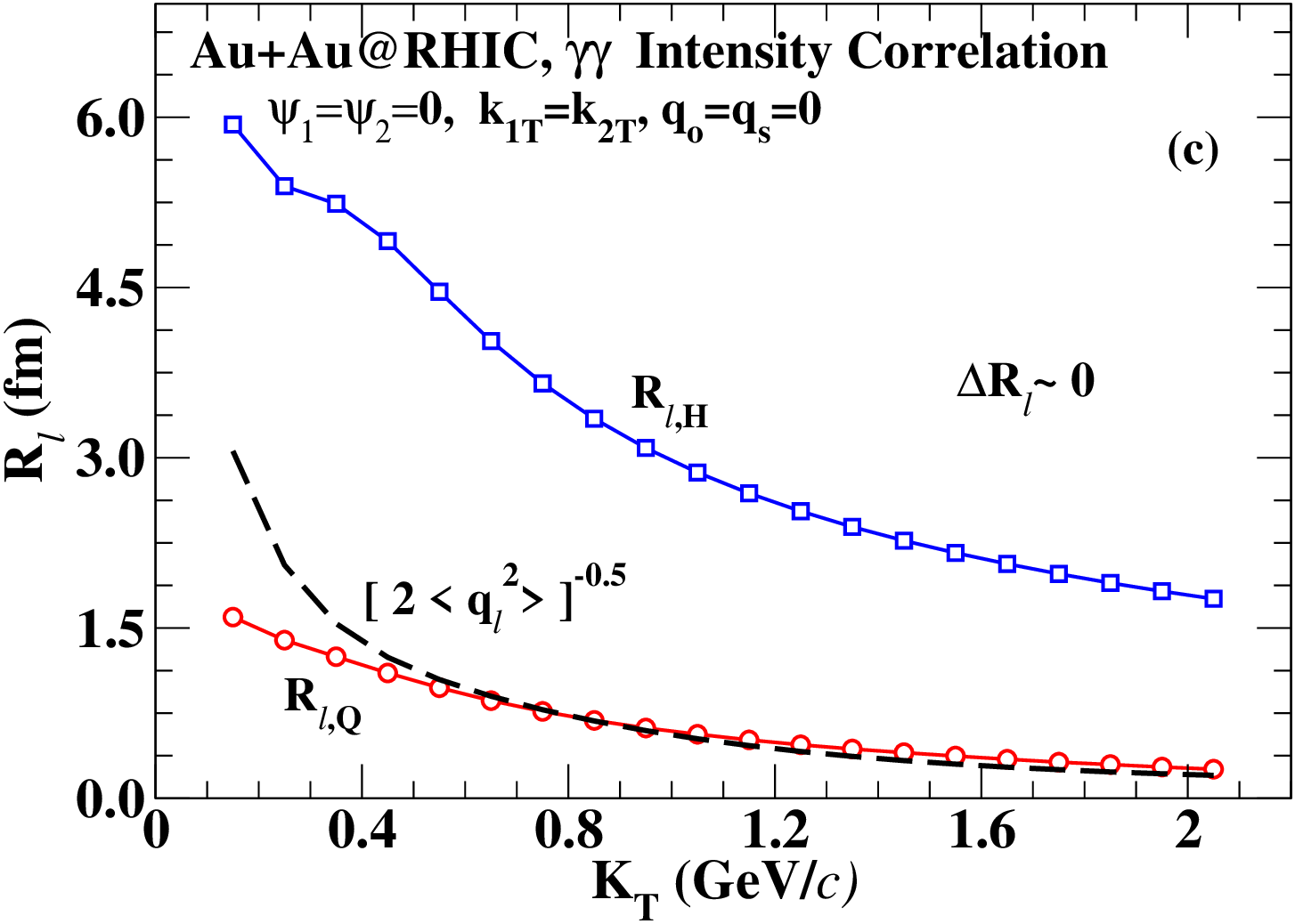}
\caption{(Colour on-line) Transverse momentum dependence 
of (a) the outward radii and temporal duration  
along with (b) the sideward and (c) the longitudinal 
radii for the hadronic and quark-matter 
sources obtained by fitting the final correlation function 
for thermal photons at RHIC energy. The radii determined
from the root mean square momentum difference for the 
correlation function are also given for a comparison~\cite{dks1}.}
\label{fig2}
\end{figure}

This is seen more clearly in Fig.~\ref{fig2} where we 
have plotted the $K_T$ dependence of $\Delta R_o$ and 
the outward, sideward, and longitudinal radii 
for the hadronic and the quark-matter sources of photons, 
obtained using the above procedure. We see that the 
outward, sideward and the longitudinal radii for the 
quark contribution 
depend weakly on the transverse momentum, indicative of 
only a mild development of the flow during that phase. The  
corresponding radii for the hadronic contribution  
show a stronger dependence on the transverse momentum, 
which is indicative of a more robust development of the 
radial flow during the hadronic phase. The duration of the 
source reveals a very interesting structure, and in fact it 
rises slightly at higher transverse momenta as the photons 
emitted during the hadronic phase benefit from the strong 
radial flow to greatly increase their transverse momenta.
%%%%%%%%%%%%%%%%%new%%%%%%%%%%%%%%%%%%%%%%%%%%%%%%%%%%%%%%%%%%%%%%%%%%%%%%
The saturation in $\Delta R_o$ towards low $K_T$ has 
its origin in the competition between radial expansion 
and the decoupling of hadronic matter as it cools down 
below the freeze-out temperature at the edges. 

The inverse root mean square momentum which, as we commented 
earlier, is a measure of the correlation radius, is seen 
to vary rapidly with $K_T$, for the outward correlation 
due to rapid variation of the competing contributions of 
the quark matter and hadronic matter phases. For the 
sideward correlation, it goes over smoothly from a value
 which is close to that for the hadronic matter at lower 
$K_T$ to that for the quark matter at higher $K_T$, as 
one would expect. (The slight difference of $1/[2\, 
\langle q_s^2 \rangle]^{1/2}$ from $R_{s,Q} \, \approx 
\, R_{s,H}$ at large $K_T$ is due to the deviation of 
the calculated correlation function  from a perfect 
Gaussian, to which it is fitted.)

%%%%%%%%%%%%%%%%%new%%%%%%%%%%%%%%%%
%%%%%%%%%%%%%%%%%%%%%%%%%%%%%%%%%%%%
\begin{figure}[tb]
\includegraphics[height=5.60cm, clip=true]{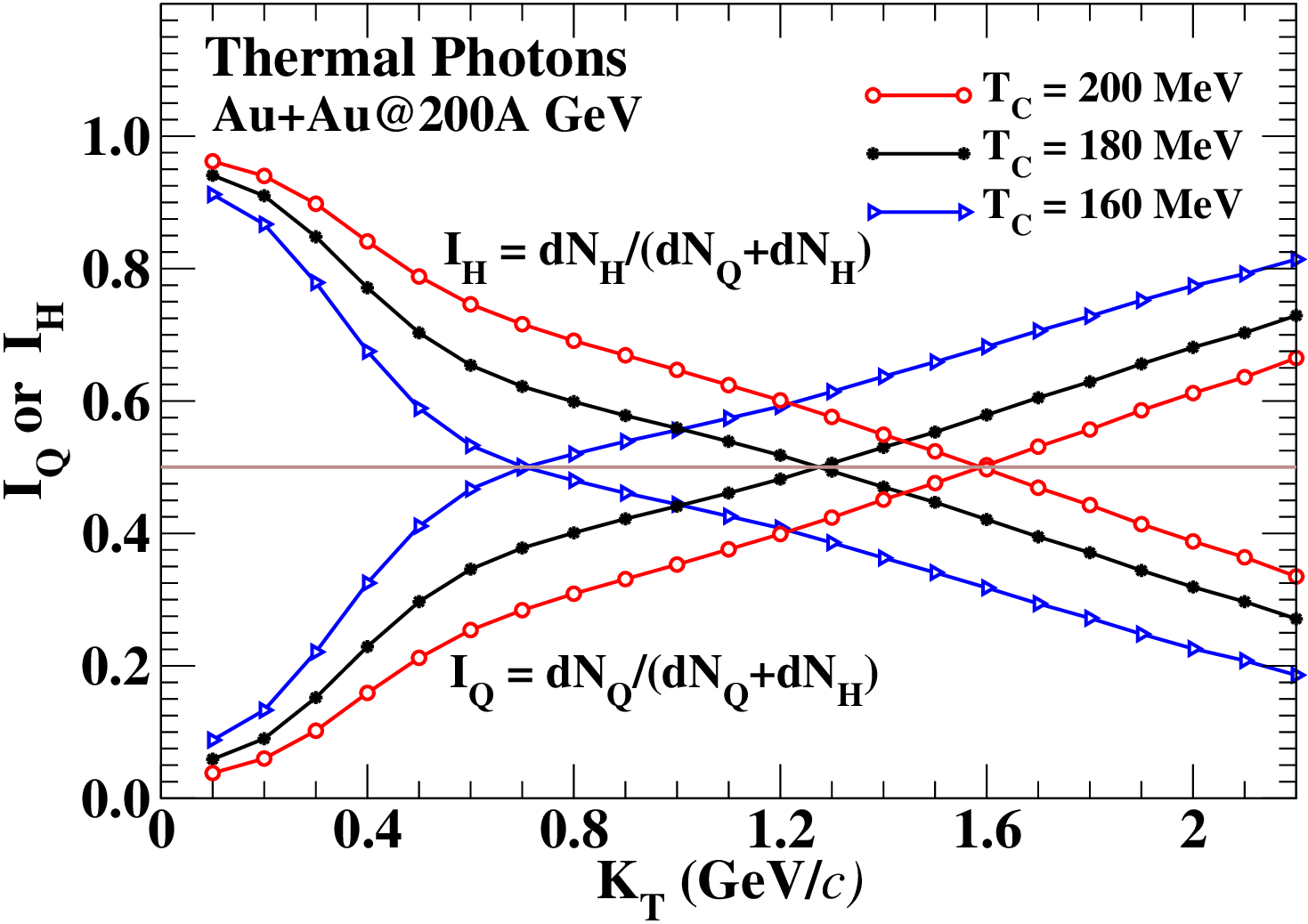}
\caption{(Colour on-line) Transverse momentum dependence 
of fraction of thermal photons from quark matter 
($\rm {I_Q}$) and hadronic matter ($\rm {I_H}$) at 
RHIC energy. The solid curves gives the results for 
$T_C$ = 180 MeV, while the dashed and dot-dashed 
curves give the results for $T_C$ equal to 160 
and 200 MeV respectively. (see text).}
\label{fig3}
\end{figure}
%%%%%%%%%%%%%%%%%%%%%%%%%%%%%%%%%%%%
\begin{figure}[tb]
\includegraphics[height=5.60cm, clip=true]{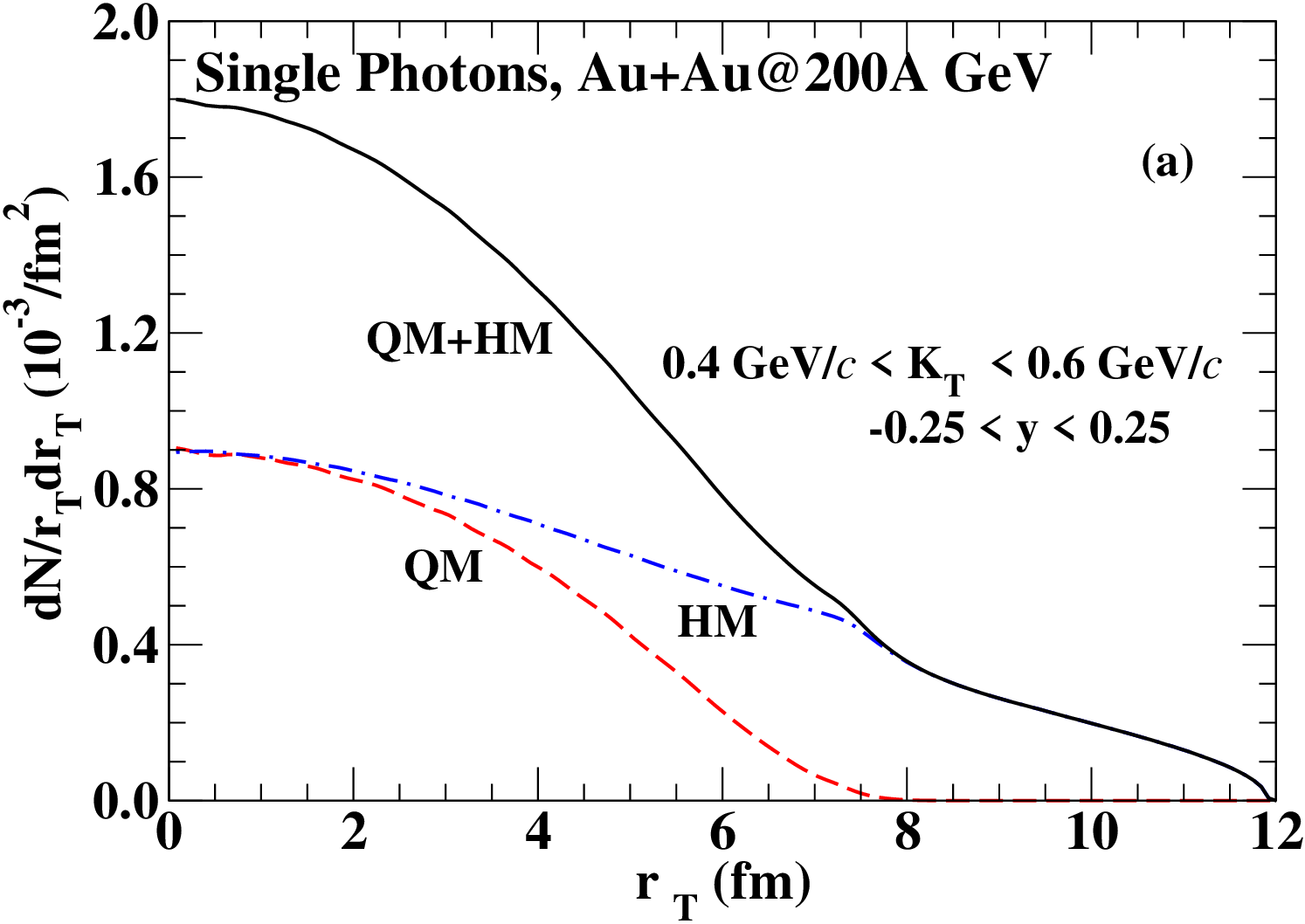}
\includegraphics[height=5.60cm, clip=true]{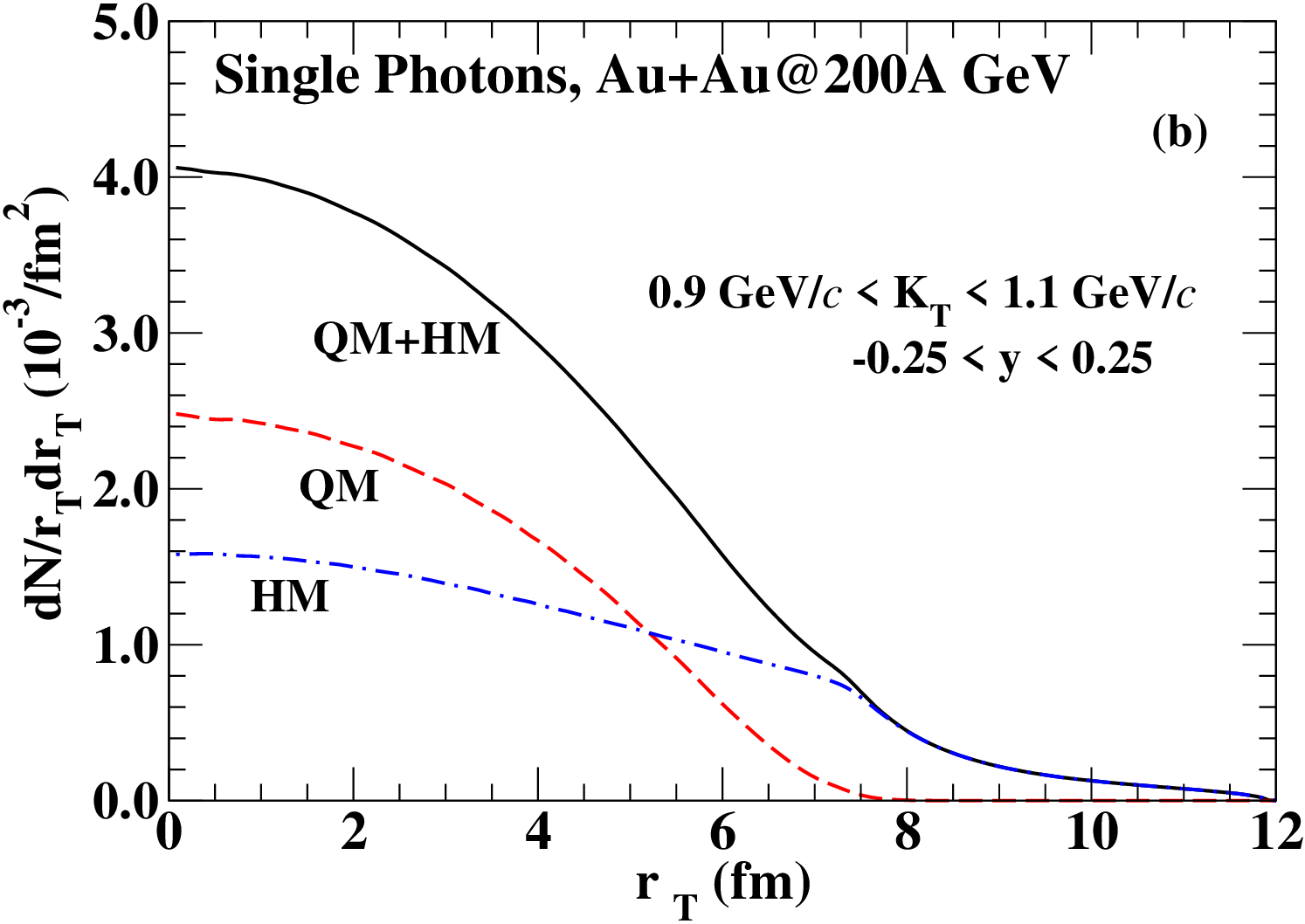}
\includegraphics[height=5.60cm, clip=true]{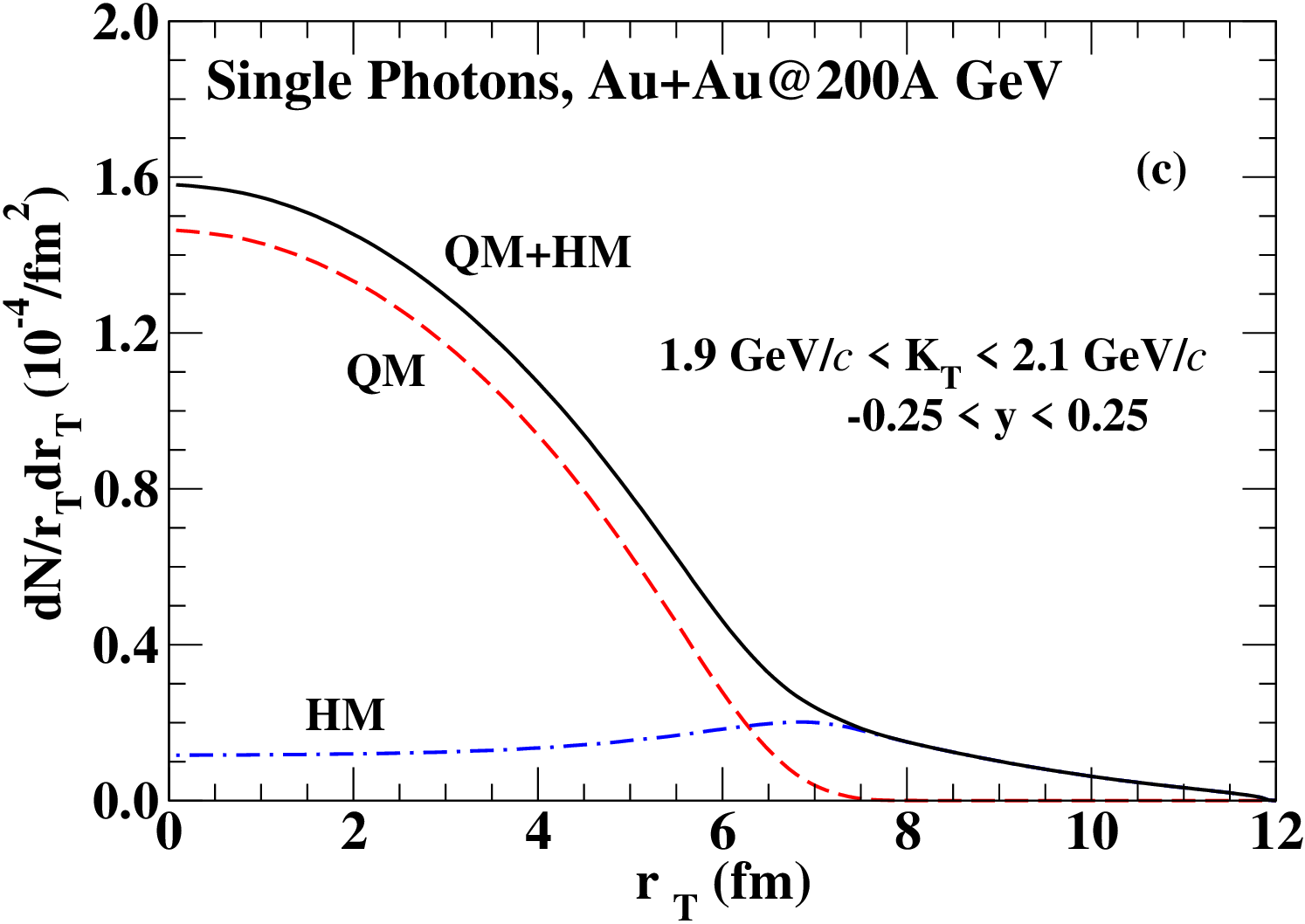}
\caption{(Colour on-line) The radial  
dependence of the source-distribution function for 
emission of photons having  transverse momenta of 
(a) 0.5, (b) 1.0, and (c) 2.0 GeV/$c$ at
RHIC energy.}
\label{fig4}
\end{figure}
%%%%%%%%%%%%%%%%%%%%%%%%%%%%%%%%%%%%
\begin{figure}[tb]
\includegraphics[height=5.60cm, clip=true]{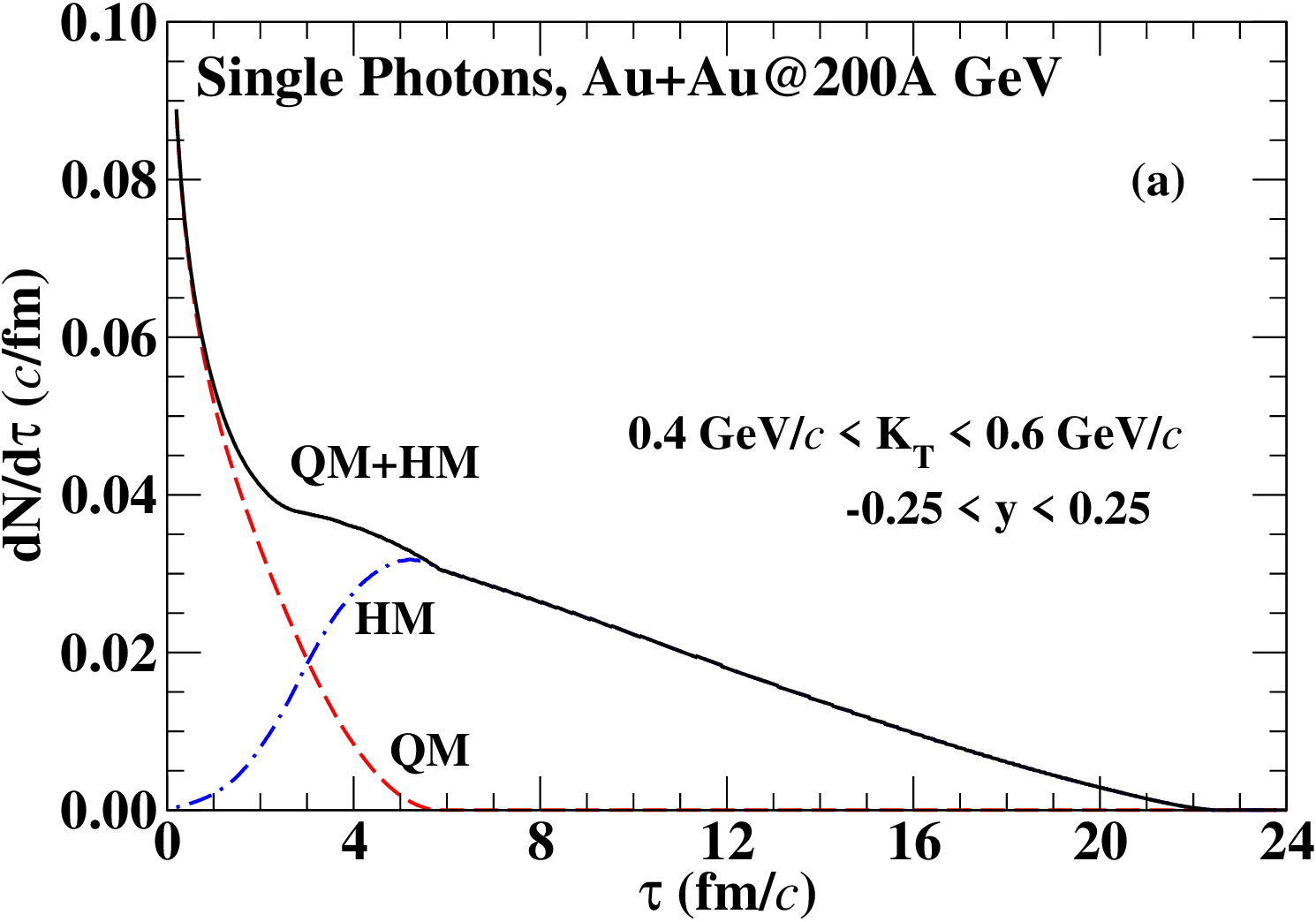}
\includegraphics[height=5.60cm, clip=true]{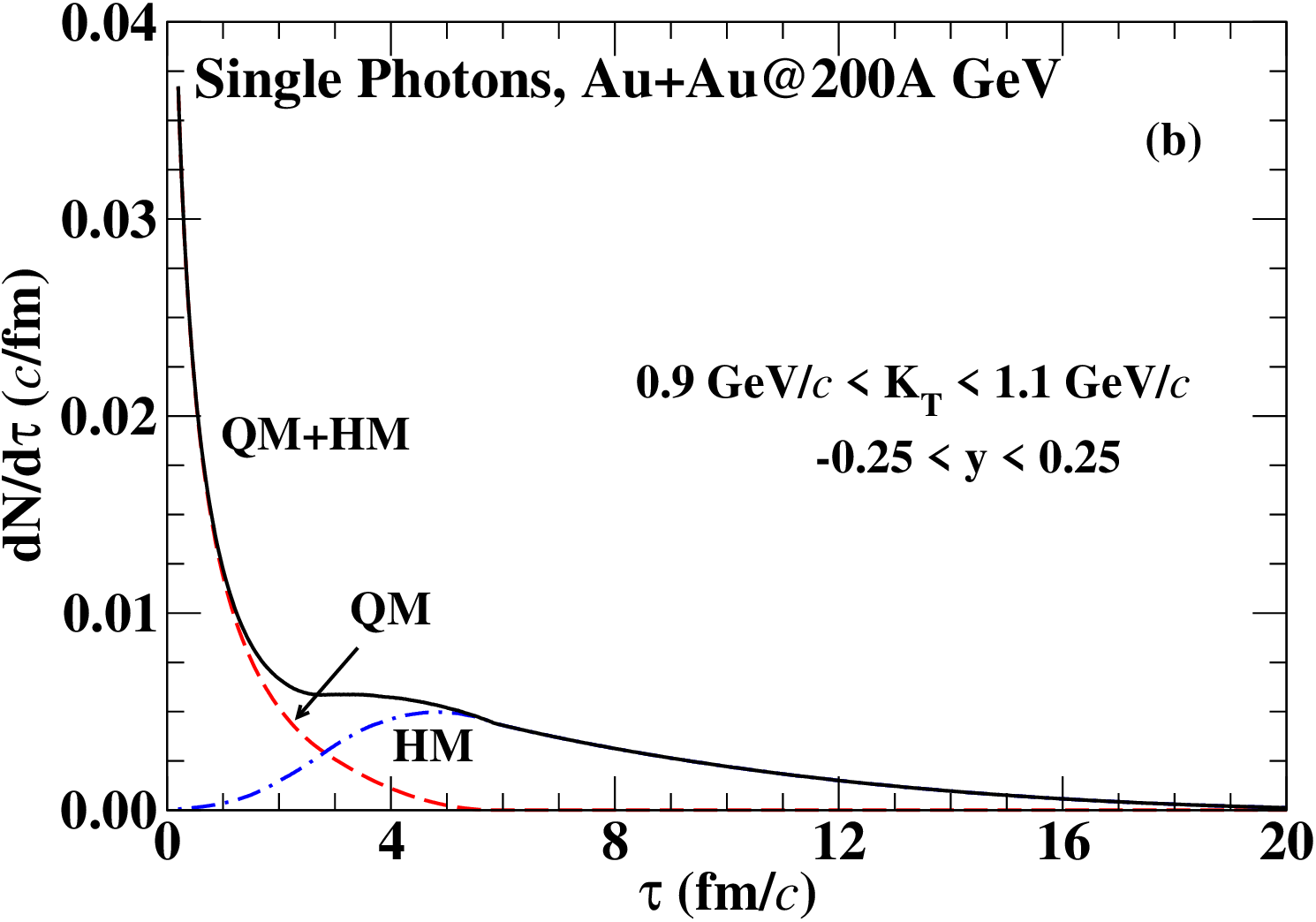}
\includegraphics[height=5.60cm, clip=true]{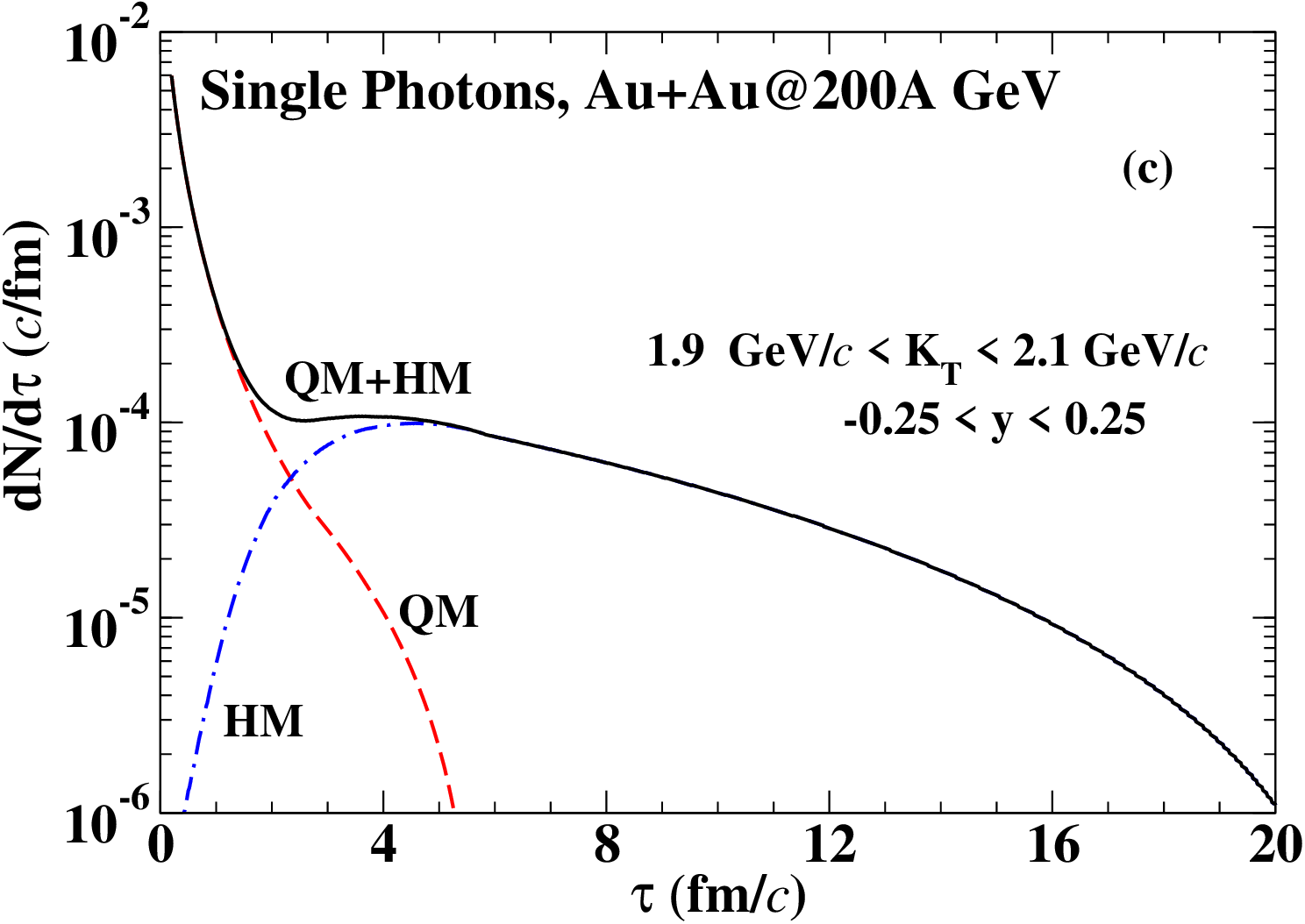}
\caption{(Colour on-line) The temporal dependence of the source
distribution function for radiation of photons having transverse 
momenta of (a) 0.5, (b) 1.0, and (c) 2.0 GeV/$c$ at RHIC energy.}
\label{fig5}
\end{figure}
%%%%%%%%%%%%%%%%%%%%%%%%%%%%%%%%%%%%
As the fractions of the quark matter and the hadronic matter 
contributions play such a pivotal role in the momentum 
dependence of overall correlation function, we study them 
next. The fractions $\rm {I_Q}$ and $\rm {I_H}$ taken from 
the calculations for the single photon spectra are shown  
in Fig.~\ref{fig3}, along with their dependence on the critical 
temperature. We shall discuss the importance of this dependence 
a little later.

In order to understand these interesting results more clearly,
we have plotted the source-functions as a function of the 
transverse distance and time for thermal photons having $K_T 
\approx$ 0.5, 1.0, and 2.0 GeV/$c$, in Fig.~\ref{fig4} and 
Fig.~\ref{fig5}. In 
all the cases we see that the radial distributions for the QGP 
as well as for  the hadronic phase are centered at $r_T$ near 
zero and that the source function for the hadronic phase extends 
considerably beyond the same for the QGP phase. This of-course is
only to be expected, due to the large transverse expansion of the 
system. We also note a good yield of thermal photons from the 
hadronic phase at larger radii. % more clearly for $K_T=$ 2 GeV.
%%%%%%%%%%%%%%%%%%%%%%new%%%%%%%%%%%%%%%%%%%%%%%
In fact, one can notice a slightly enhanced emission at larger radii 
compared to that from the central region for $K_T \, = \, 2$ GeV/$c$.
%%%%%%%%%%%%%%%%%%%%%%new%%%%%%%%%%%%%%%%%%%%%%%
This, as we noted earlier, arises due to the large kick received 
by the photons due to the radial expansion, leading to the 
blue-shift of their transverse momenta. The relative importance 
of the two contributions will depend on $K_T$ and the transition 
temperature and lead to a rich structure
in the resulting correlation function. However, note that the 
second term in the source of photons from the hadronic matter 
becomes important at larger $K_T$, where the contribution of the 
hadronic phase to photons is rather small, and thus it could be 
difficult to detect this effect.

Even a more valuable insight is provided by the temporal structure 
(Fig.~\ref{fig5}) of the emission of photons from the QGP and the hadronic 
sources; the latter emerging after a lapse of some time, which 
the system spends in the QGP phase. The duration over which the 
photons from the hadronic phase are emitted is quite large and 
essentially controls the parameter $\Delta R_o$. It is worth 
recalling that the temporal structure of the source function seen 
here is qualitatively similar to the one seen in nucleus-nucleus
collisions at cyclotron energies~\cite{yves}.

%%%%%%%%%%%%%%%%%%%%%%%%%%%%
\begin{figure}[tb]
\centerline{\includegraphics[width=8.50cm, clip=true]{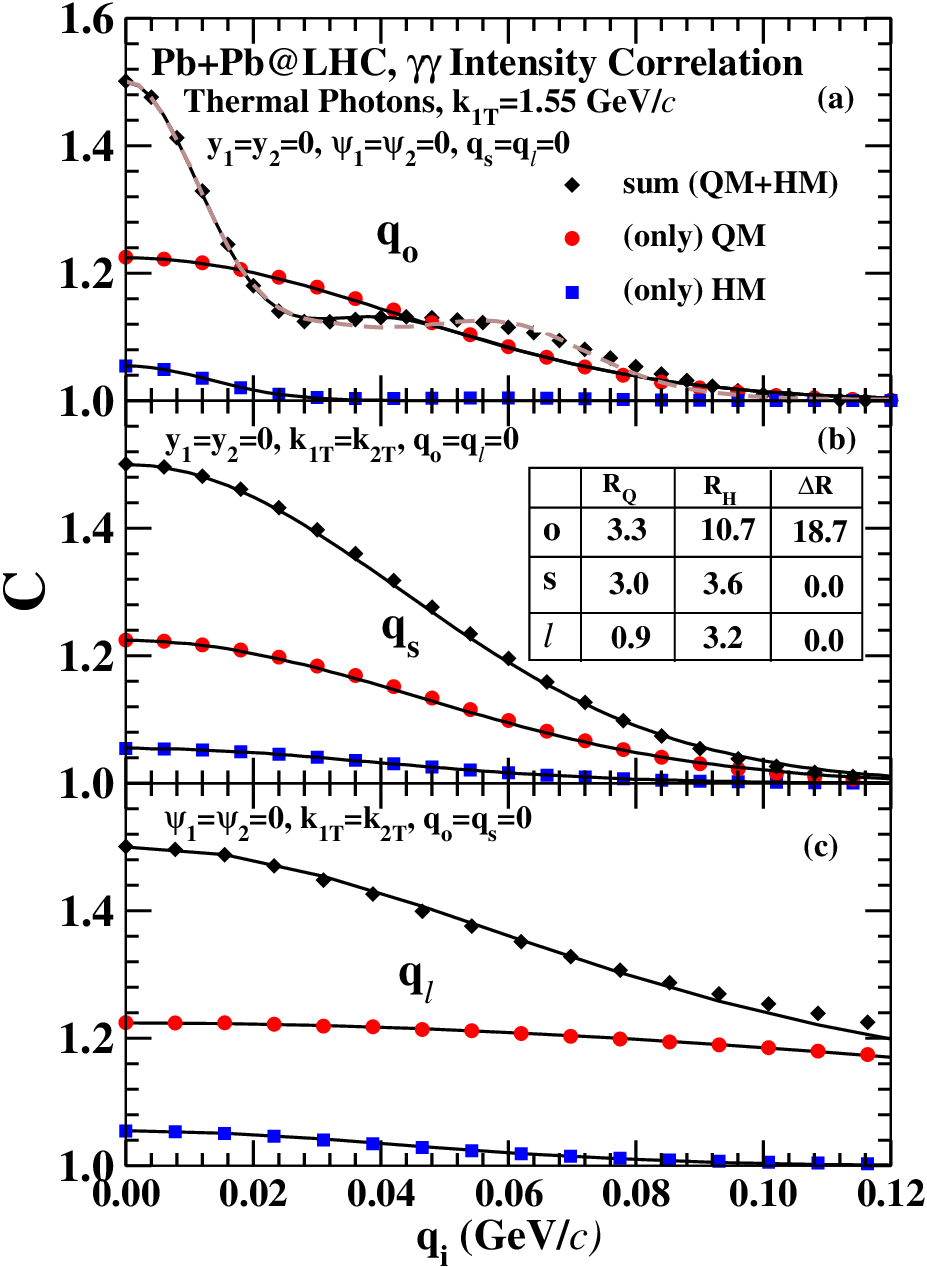}}
\caption{(Colour on-line) Same as Fig.~\ref{fig1} for  Pb+Pb 
at LHC.}
\label{fig6}
\end{figure}
%%%%%%%%%%%%%%%%%%%%%%%%%%%%%%%%%%%%%%%%%%%%%%%%%%%%5
\begin{figure}[tb]
\includegraphics[height=5.60cm, clip=true]{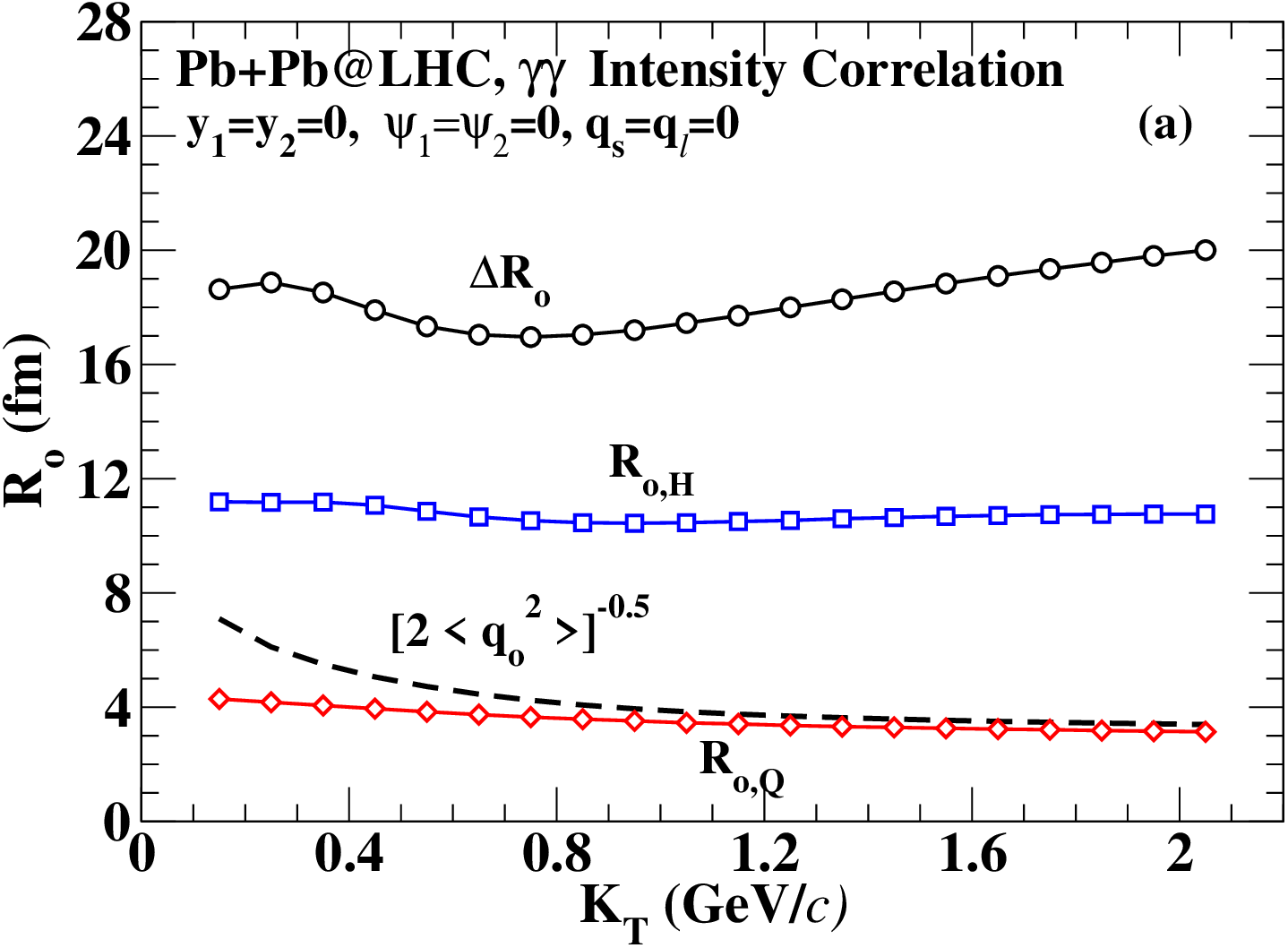}
\includegraphics[height=5.60cm, clip=true]{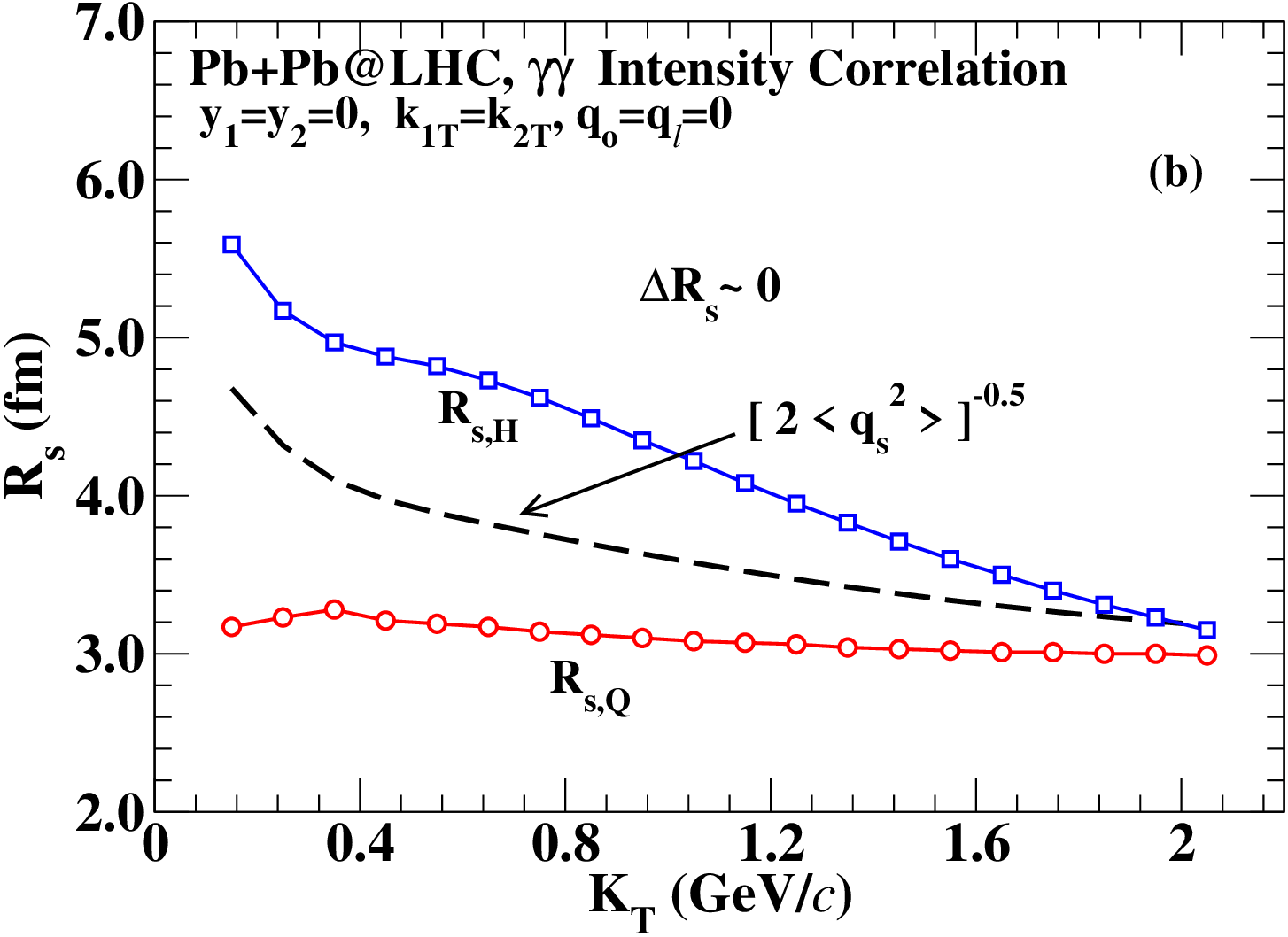}
\includegraphics[height=5.60cm, clip=true]{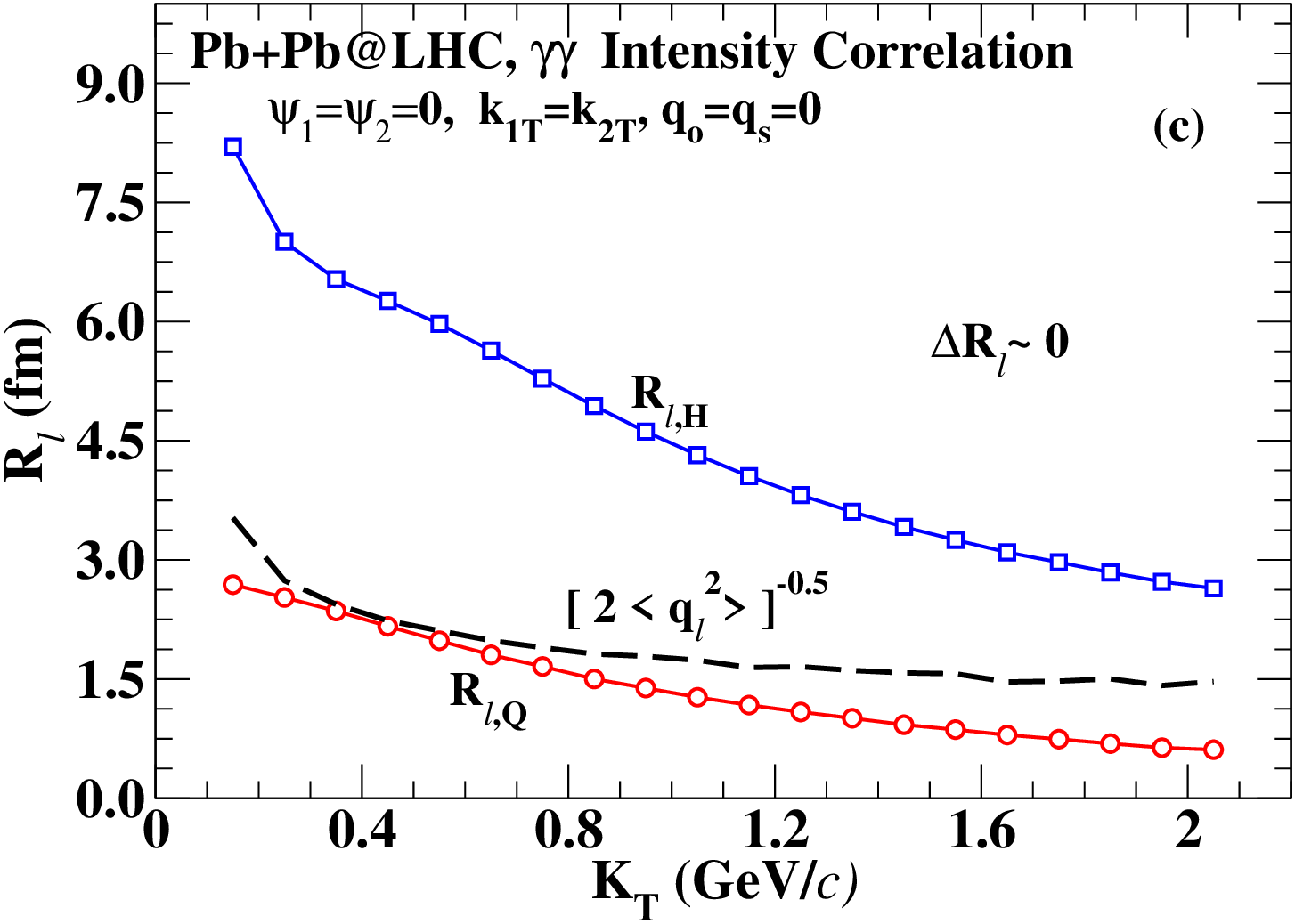}
\caption{(Colour on-line) Transverse momentum dependence 
of (a) the outward radii and temporal duration 
along with (b) the sideward, and (c) the longitudinal 
radii for the hadronic and quark-matter 
sources obtained by fitting the final correlation function 
for thermal photons at LHC energy. The radii determined
from the root mean square momentum difference for the 
correlation function is also given for a comparison~\cite{dks1}.}
\label{fig7}
\end{figure}
%%%%%%%%%%%%%%%%%%%%%%%%%%%%%%%%%%%%%%%%%%%%%%%%%%%%%%
\begin{figure}[tb]
\includegraphics[height=5.60cm, clip=true]{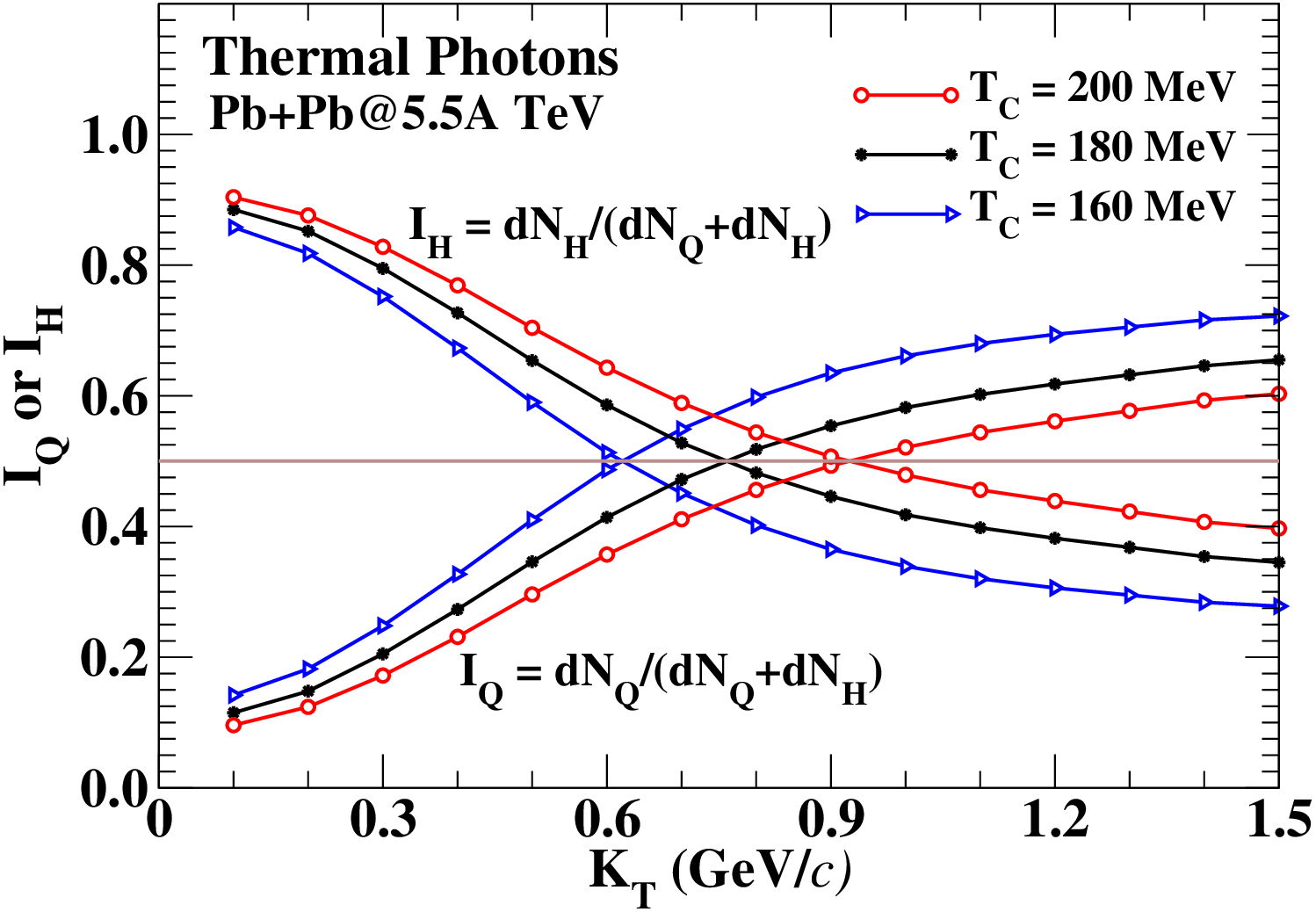}
\caption{(Colour on-line) Transverse momentum dependence 
of fraction of thermal photons from quark matter 
($\rm {I_Q}$) and hadronic matter ($\rm {I_H}$) at 
LHC energy. The solid curves gives the results for 
$T_C$ =180 MeV, while the dashed and dot-dashed curves 
give the results for $T_C$ equal to 160 and 200 MeV 
respectively.}
\label{fig8}
\end{figure}
%%%%%%%%%%%%%%%%%%%%%%%%%%%%%%%%%%%%%%%%%%

\section{Results for LHC energies}
The Large Hadron Collider will study collisions of lead nuclei at 
unprecedented energies of 5500 GeV/A. It is expected that the 
initial temperature likely to be attained reached in such collisions would be much 
higher than that at RHIC energies. This would provide a golden 
opportunity to study the properties of the QGP and the dynamics 
of its evolution. A larger initial temperature would lead to a 
larger duration of the interacting system, which in turn will 
provide an ample opportunity to the mechanism of expansion to 
develop. We can thus expect to put our models about the evolution 
of the interacting system to a very rigorous test.

In Fig.~\ref{fig6} we show our results  for the intensity 
correlation of thermal photons having $K_T \approx 1.5$ 
GeV/$c$ produced at LHC. We see, as before, an  interference 
of the photons from the hadronic and QGP phases of the system.  
We add that the fit to the calculated values for the outward 
correlation function can be improved considerably (see dashed 
curve) by adding one more Gaussian term, centered at $q_0 
\approx 0.06$ GeV/$c$ in the source term for the hadronic 
matter, or by approximating $\rho_{o,H} \approx \sqrt{2(C_H-1)}$, 
where $C_H$ is the corresponding correlation function obtained 
numerically. We are trying to understand this observation.

The results for $C(q_s)$ and $C(q_\ell)$ are similar in nature 
to those found at RHIC energy. We note that now the temporal 
separation of the two sources is about 19 fm/$c$ (see inset 
Fig.~\ref{fig6}).  

%%%%%%%%%%%%%%%%%%%%new%%%%%%%%%%%%%%%%%%%%%%%%%%%%%%%%%%%%%%%%%%%%%%%%%%%
Next we discuss the $K_T$ dependence of the correlation radii 
(Fig.~\ref{fig7}). We see a behaviour which is qualitatively 
similar to the one obtained earlier, though all the correlation 
radii are larger, specially $\Delta R_o$ which gives the duration 
of the source. We do emphasize here that $1/[2\, \langle q_i^2 
\rangle]^{1/2}$ is closer to $R_{i,Q}$ at LHC due to the dominance of 
quark matter contribution even at modest $K_T$.

In Fig~\ref{fig8}, we have plotted the fractions of momentum 
dependence of quark and hadron contribution at LHC energies, 
at three transition temperature $T_C\,( = \, 160, \, 180, \, 
\rm{and} \, 200\, MeV)$. Comparing these results with those 
of Fig.~\ref{fig3}, we see once again that decreasing $T_C$ 
increases the fraction of photons coming from the quark matter. 
We also note that the transverse momentum where the quark and 
hadronic contributions become equal is quite sensitive to the 
transition temperature $T_C$. Realizing that, in an ideal 
situation we would be able to decompose the outward, sideward, 
and longitudinal correlations into two sources and that their 
intercepts on the y-axis will give $\rm {I_Q}$ and $\rm {I_H}$, 
this opens up the tantalizing possibility of determination of the 
transition temperature, the two fractions becoming equal at $K_T$.

\begin{figure}[tb]
\includegraphics[height=5.60cm, clip=true]{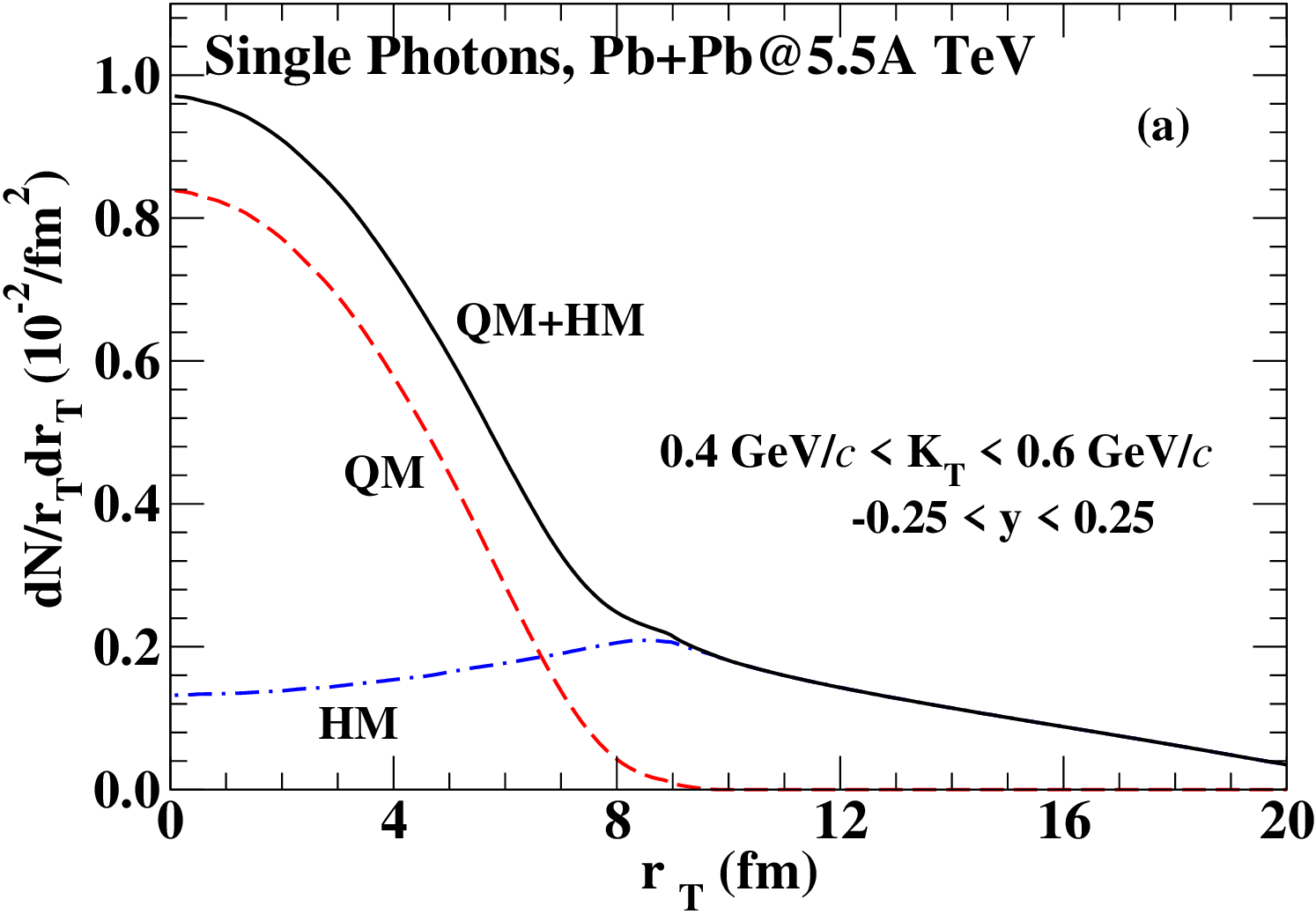}
\includegraphics[height=5.60cm, clip=true]{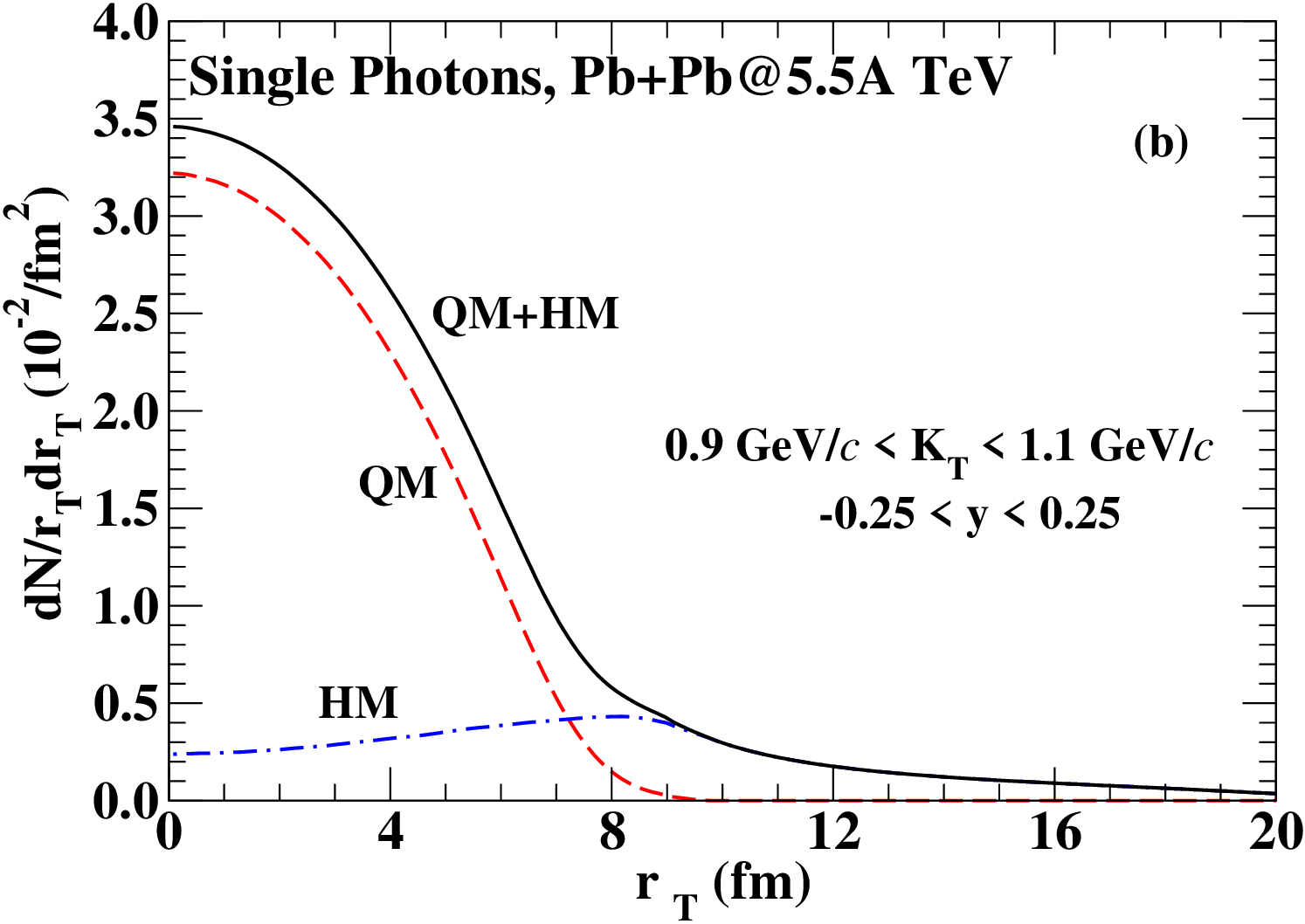}
\includegraphics[height=5.60cm, clip=true]{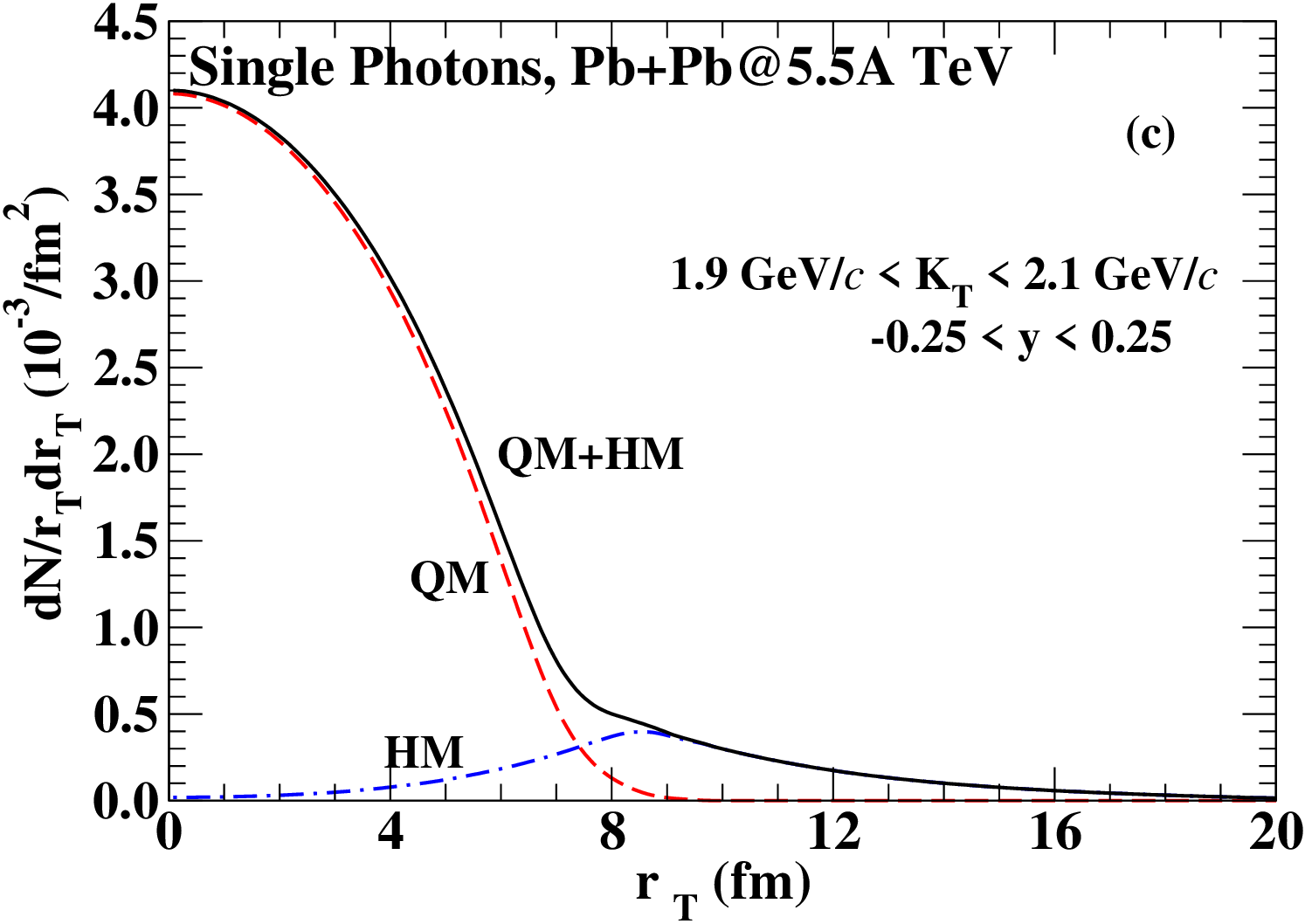}
\caption{(Colour on-line) The radial 
dependence of the source-distribution function
for emission of photons having  transverse momenta 
(a) 0.5, (b) 1.0, and (c) 2.0 GeV/$c$ at
LHC energiy.}
\label{fig9}
\end{figure}

\begin{figure}[tb]
\includegraphics[height=5.60cm, clip=true]{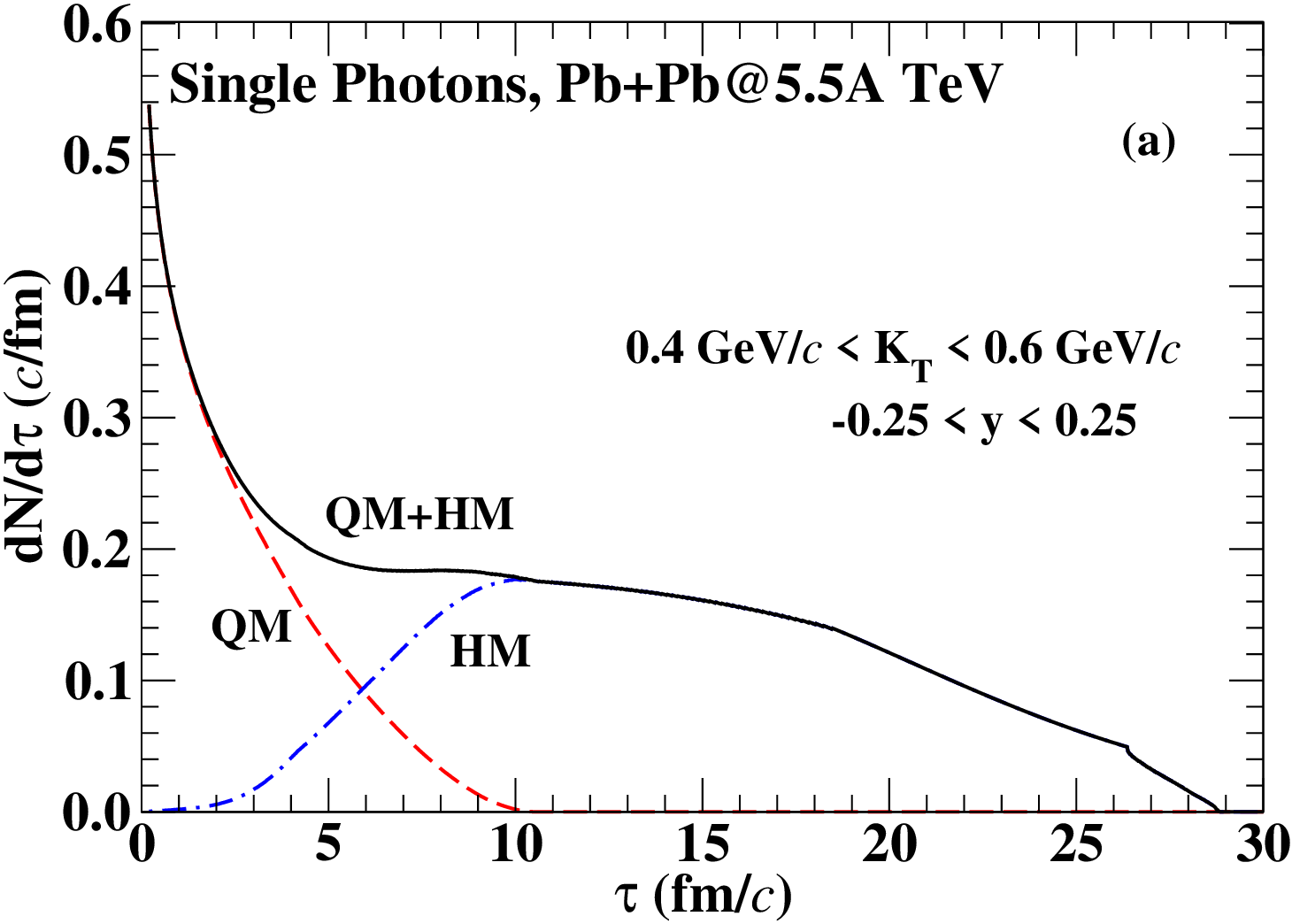}
\includegraphics[height=5.60cm, clip=true]{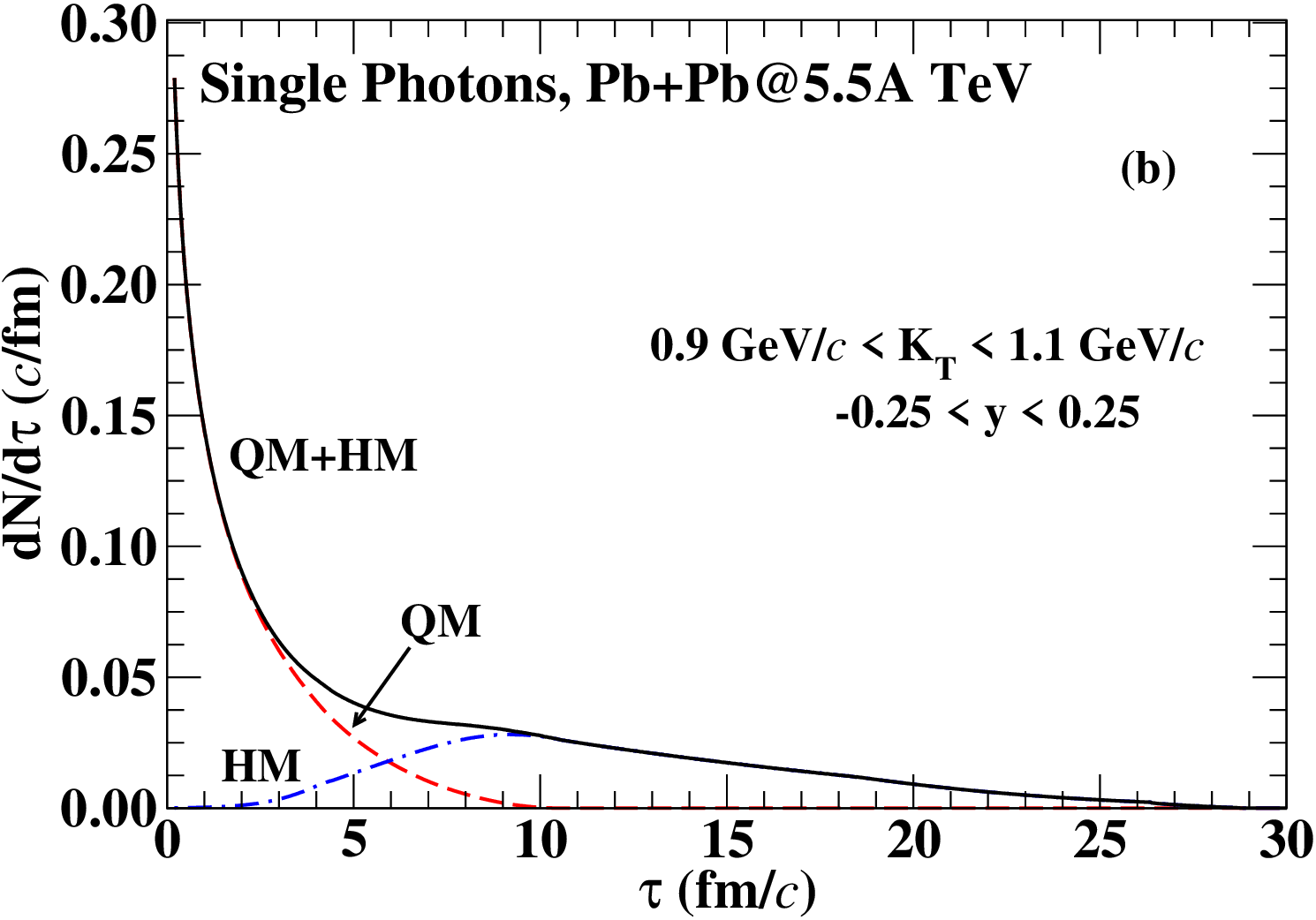}
\includegraphics[height=5.60cm, clip=true]{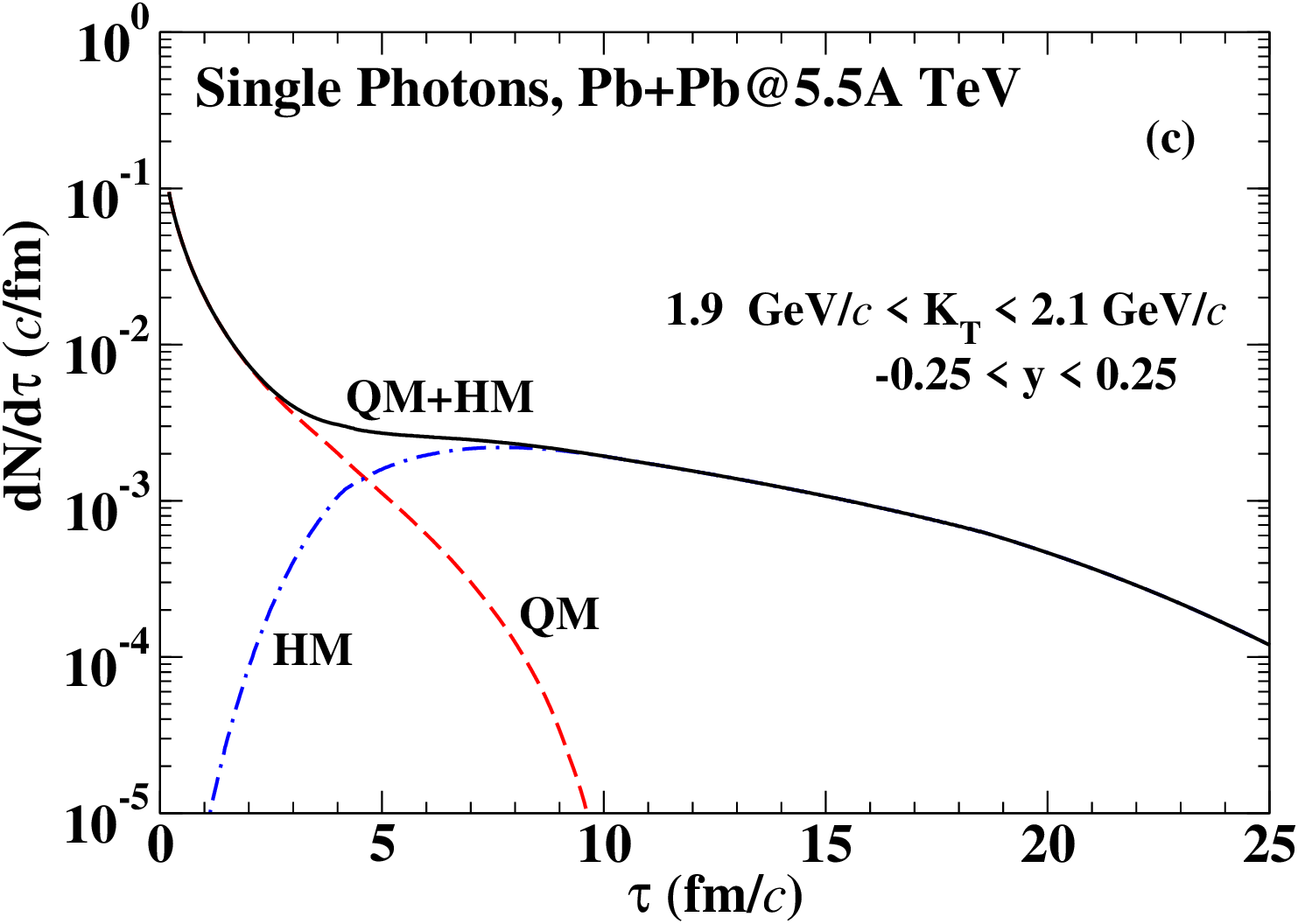}
\caption{(Colour on-line) The temporal dependence of the source
distribution function for radiation of photons having transverse 
momenta of (a) 0.5, (b) 1.0, and (c) 2.0 GeV/$c$ at LHC energy.}
\label{fig10}
\end{figure}

We next discuss the spatial (Fig.~\ref{fig9}) and temporal  
(Fig.~\ref{fig10}) distribution of source function for photons 
having $K_T \, = \, 0.5, \, 1.0, \, \rm{and} \, 2.0$ GeV/$c$ 
produced in central collision of Pb nuclei at LHC energy. Even though 
we note that the source distributions at LHC are qualitatively 
similar to those at RHIC, the extension of the hadronic sources 
beyond those for the quark sources (Fig.~\ref{fig9}) and 
enhanced production at larger $r_T$ from hadronic sources 
are amplified considerably here. The dying out of the quark 
source and a delayed build-up of hadronic source in time 
(Fig.~\ref{fig10}) are seen to emerge very clearly.
%%%%%%%%%%%%%%%%%%%%new%%%%%%%%%%%%%%%%%%%%%%%%%%%%%%%%%%%%%%%%%%%%%%%%%%%

% fig.7 l-out- k-T
%       l-side-kt
% fig.8 I_Q
%fig.9 source 1_r,1_t
%fig.10 source 2_r,2_t
%%%%%%%%%%%%%%%%%%%%%%%%%%%%%%%%%%%%
%%%%%%%%%%%%%%%%%%%%%%%%%%%%%%%%%%%%

\section{Sensitivity to $\tau_0$ and $T_C$}
We have seen that the final correlation is decided by 
the relative contributions from the quark-matter which 
occupies a smaller volume and is shorter lived, and those from 
the hadronic matter, which occupies to a larger volume and 
lives longer. %if the evolution admits a mixed phase.

\begin{figure}[tb]
\centerline{\includegraphics[height=8.50cm, clip=true]{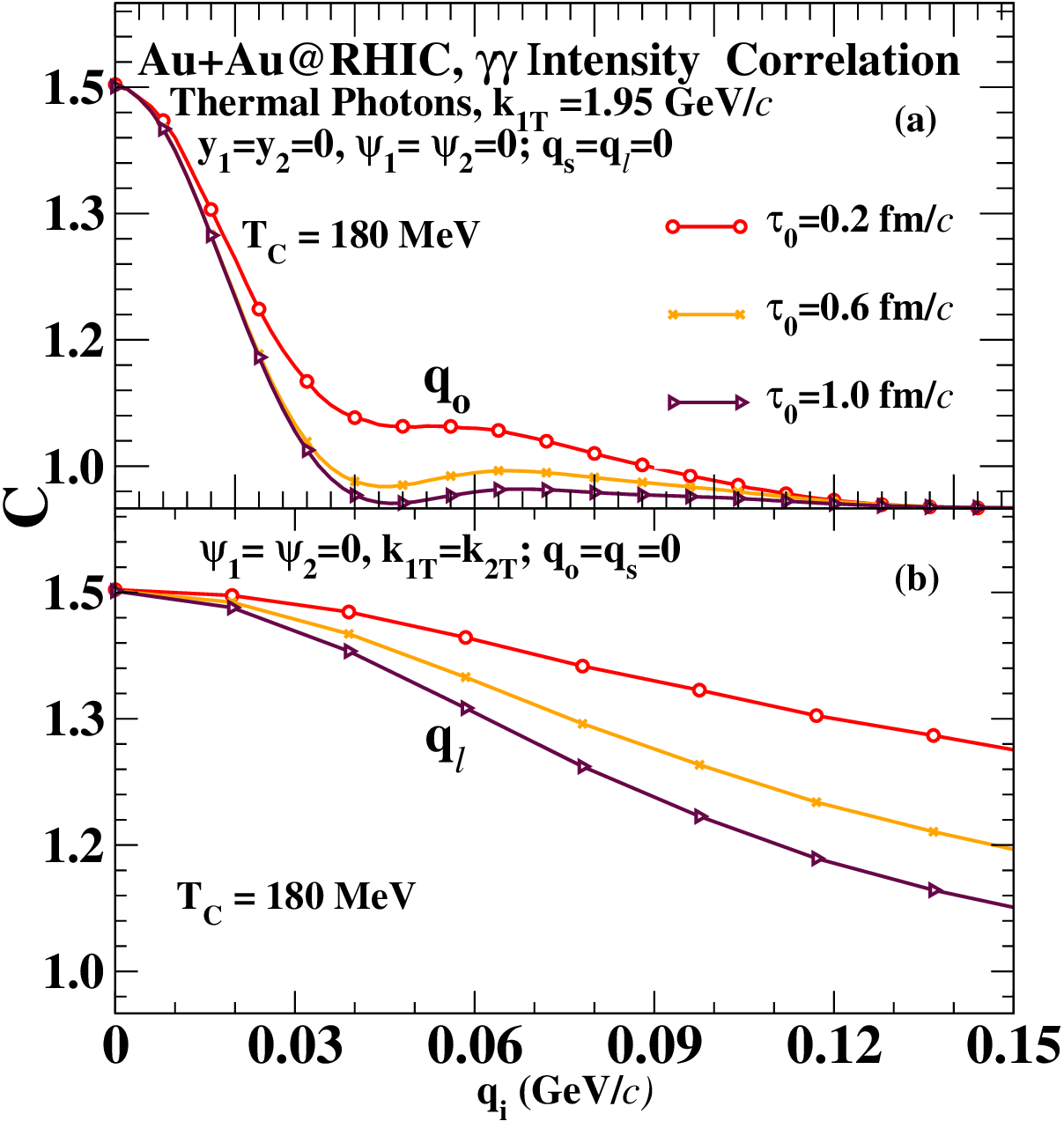}}
\caption{(Colour on-line) $\tau_0$ dependence of (a) the outward 
and (b) the longitudinal correlation function at RHIC.}
\label{fig11}
\end{figure}

If the system thermalizes quickly, the initial temperature 
would be large. One can test the sensitivity of the results 
to the formation time of the plasma, by considering systems 
with identical entropy, but having varying formation times 
($\tau_0$). The fractional contribution of the quark matter, 
($\rm {I_Q}$) will increase with decreasing $\tau_0$. In 
Fig.~\ref{fig11} we have shown the $\tau_0$ dependence of the 
outward and longitudinal correlations at RHIC. The sideward 
correlation function did not show any perceptible change due 
to the variation of $\tau_0$ and is not shown  here. 
%%%%%%%%%%%%%%%%%%%%%%%%%%%
%%%%%%%%%%%%%%%%%%%%%%%%%%%
%%%%%%%%%%%%%%%%%%%%%%%%%%%%
%%%%%%%%%%%%%%%%%%%%%%%%%%%%
\begin{figure}[tb]
\centerline{\includegraphics[height=5.60cm, clip=true]{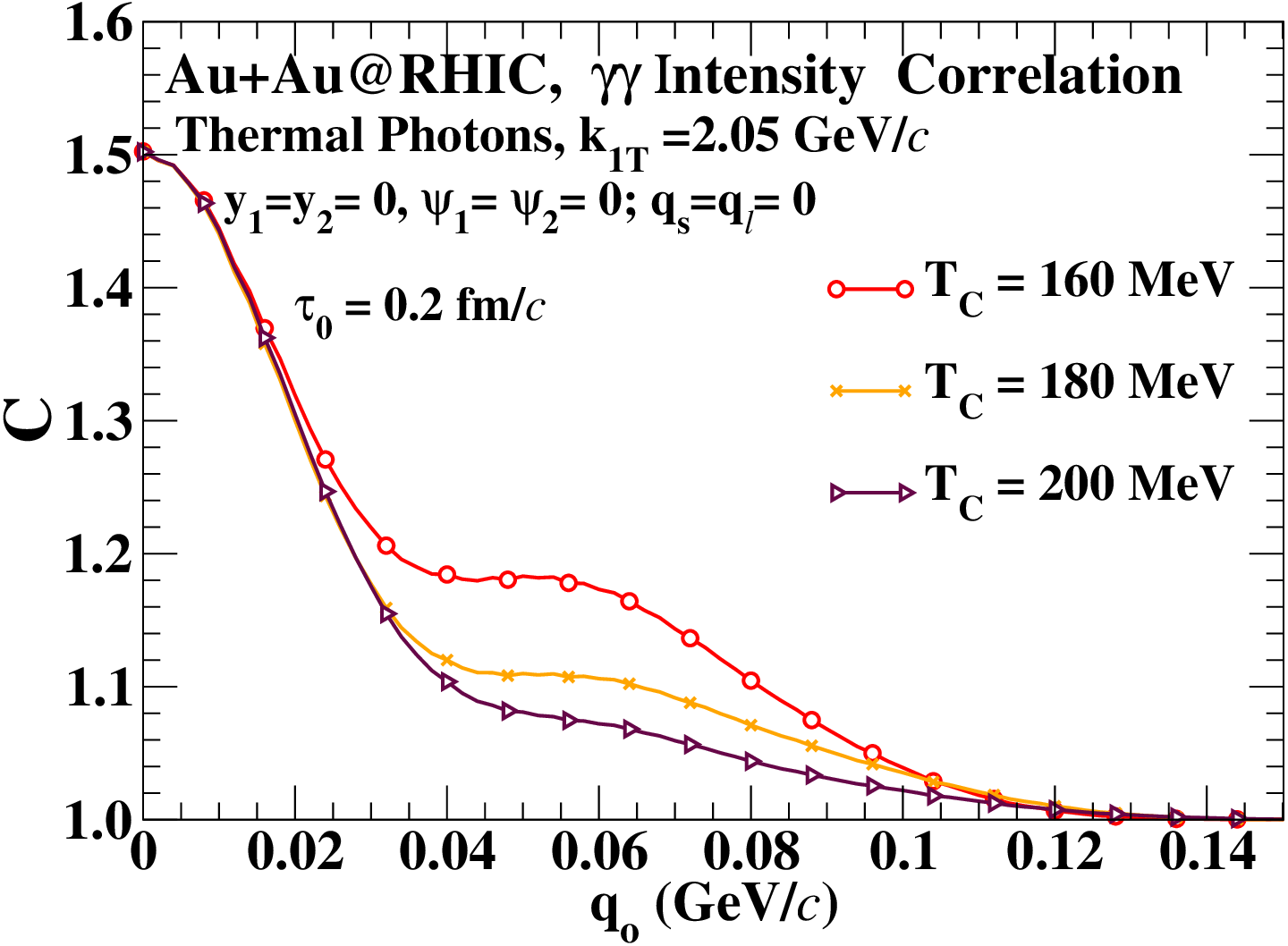}}
\caption{(Colour on-line) $T_C$ dependence of the outward 
correlation function at RHIC.}
\label{fig12}
\end{figure}
%%%%%%%%%%%%%%%%%%%%%%%%%%%%%%%%%
We must add here that choosing a large $\tau_0$ would necessitate the
inclusion of the pre-equilibrium contribution to the photons
~\cite{bms_phot} which must surely be there, at least at 
larger $K_T$. The jet-conversion mechanism~\cite{fms_phot} is 
also likely to contribute at larger $K_T$. These contributions, 
per-force have their origin in the deconfined matter 
and their spatial distribution is not likely to be very different 
from the early stages of the initial distributions assumed here. 
Their contributions would increase $\rm {I_Q}$ in these 
studies, which in turn could mimic an effectively smaller $\tau_0$. 
This would still be useful, as it will amount to getting an 
effective $\tau_0$ after which the hydrodynamics can be applied.

However these discussions open the door to another interesting 
and potentially powerful observation, perhaps,  with  a far 
reaching implication. We have already noted the sensitivity 
of the fractions of the contributions of the  quark-matter 
($\rm {I_Q}$) and the hadronic matter ($\rm {I_H}$) to the 
transition temperature at both RHIC and LHC (see Figs.~\ref{fig3} 
and~\ref{fig8}). In Fig.~\ref{fig12} we show our results for the 
sensitivity of the outward correlation at RHIC to the transition 
temperature. Recalling that an increase in the transition 
temperature leads to a decrease in $\rm {I_Q}$, and that the 
quark matter contribution has a smaller $R_o$ the change in the 
interference pattern seen at larger $q_o$ is easily understood.

It is felt that the results shown here should be valid 
whenever the correlations arise from contributions from two 
sources separated in space and time (see e.g., 
Ref.~\cite{yves}). The contribution of the two sources to the 
correlation function also provides a natural explanation for the 
failure of the earlier studies~\cite{dks1} to find a simple 
Gaussian parametrization for it.

\section{Summary}
To summarize, the rich structure of the sideward, outward, and 
longitudinal correlation functions for intensity interferometry 
of thermal photons at RHIC and LHC energies is calculated. The 
correlation functions are marked by a very distinctive interference 
between the photons from the quark and the hadronic matter, which 
is most clearly visible in the outward  correlation. We have 
calculated the transverse momentum dependence of the correlation 
radii and the duration of the emission and tried to understand 
their behaviour by calculating the spatial and the temporal 
distribution of the sources and their contributions. The study 
has thrown open an interesting possibility of determination of 
transition temperature and the formation time of the plasma.
Finally, we would like to add that even though several  earlier studies 
have also  talked of the two sources of photons (quark and hadronic 
matter) contributing to the correlation function, the present 
work, as far as we know, is the first attempt to study their 
interference in relativistic heavy ion collisions.
%This helps us provide a quantitative method to determine the lifetime of the system, the relative strength of the two contributions, and the thermalization time of the system.

\section*{Acknowledgment}
We thank E. Frodermann, U. Heinz, G. Martinez, and Y. Schutz 
for  very useful and informative discussions.

\end{document}